\documentclass[12pt]{article}
\usepackage{epsfig}
\usepackage{pstricks,rotating, graphicx}
\usepackage{a4}
\usepackage{latexsym}
\usepackage{cite}

\usepackage{color}
\usepackage{colordvi}
\usepackage[hang]{subfigure}

\graphicspath{{Pics/}}
\DeclareGraphicsExtensions{.eps,.ps}

\usepackage[width=16.5cm, left=2.2cm,top=2.5cm,height=24.cm]{geometry}

\usepackage{pslatex}
\usepackage{txfonts}
\usepackage[latin1]{inputenc}
\usepackage[T1]{fontenc}

\def\one{{\rm 1\!\!l}}

\def\mt{{m_t}}
\def\mts{{m_t^2}}

\def\shat{{\hat s}}
\def\muf{{\mu^{}_f}}
\def\mufs{{\mu^{\,2}_f}}
\def\mur{{\mu^{}_r}}
\def\murs{{\mu^{\,2}_r}}

\def\msbar{{$\overline{\mbox{MS}}$}}
\def\mbar{\overline{m}}
\def\mmu{m(\mur)}
\def\mm{m(m)}
\def\lm{{L_{\mmu} } }

\def\Pqq#1{P_{qq}^{(#1)}}
\def\Pgq#1{P_{gq}^{(#1)}}
\def\Pqg#1{P_{qg}^{(#1)}}
\def\Pgg#1{P_{gg}^{(#1)}}

\def\b#1{\beta_{#1}}
\def\z#1{\zeta_{#1}}

\def\fqqn{f_{q\bar{q}}^{(0)}}
\def\fgqn{f_{gq}^{(0)}}
\def\fggn{f_{gg}^{(0)}}

\def\fqqon{f_{q\bar{q}}^{(10)}}
\def\fqqoo{f_{q\bar{q}}^{(11)}}
\def\fqqtn{f_{q\bar{q}}^{(20)}}
\def\fqqto{f_{q\bar{q}}^{(21)}}
\def\fqqtt{f_{q\bar{q}}^{(22)}}

\def\fgqon{f_{gq}^{(10)}}
\def\fgqoo{f_{gq}^{(11)}}
\def\fgqtn{f_{gq}^{(20)}}
\def\fgqto{f_{gq}^{(21)}}
\def\fgqtt{f_{gq}^{(22)}}

\def\fggon{f_{gg}^{(10)}}
\def\fggoo{f_{gg}^{(11)}}
\def\fggtn{f_{gg}^{(20)}}
\def\fggto{f_{gg}^{(21)}}
\def\fggtt{f_{gg}^{(22)}}

\newcommand{\GeV}{\ensuremath{\,\mathrm{GeV}}}
\newcommand{\TeV}{\ensuremath{\,\mathrm{TeV}}}

\newcommand{\pb}{\ensuremath{\,\mathrm{pb}}}

\newcommand{\gsim}{\raisebox{-0.07cm}{$\:\:\stackrel{>}{{\scriptstyle \sim}}\:\: $} }

\newcommand{\mbarmeq}{\raisebox{-0.07cm}{$\:\:\stackrel{m_0=\mbar}{=}\:\:$} }

\begin{document}

\begin{titlepage}
\noindent
DESY 09-097 \\
HU-EP-09/27 \\
SFB/CPP-09-55 \\
June 2009 \\
\vspace{1.3cm}
\begin{center}
\Large{\bf 
Measuring the running top-quark mass
}\\
\vspace{1.5cm}
\large
U. Langenfeld$^{\, a}$, S. Moch$^{\, a}$ and P. Uwer$^{\, b}$\\
\vspace{1.2cm}
\normalsize
{\it $^a$Deutsches Elektronensynchrotron DESY \\
\vspace{0.1cm}
Platanenallee 6, D--15738 Zeuthen, Germany}\\
\vspace{0.5cm}
{\it $^b$Humboldt-Universit\"at zu Berlin, Institut f\"ur Physik \\
\vspace{0.1cm}
Newtonstra\ss e 15, D--12489 Berlin, Germany}\\
\vspace{1.8cm}

\large
{\bf Abstract}
\vspace{-0.2cm}
\end{center}
We present the first direct determination of the running top-quark mass based 
on the total cross section of top-quark pair-production as measured at the Tevatron.
Our theory prediction for the cross section includes various next-to-next-to-leading order QCD contributions, 
in particular all logarithmically enhanced terms near threshold, the Coulomb corrections at two loops 
and all explicitly scale dependent terms at NNLO accuracy. 
The result allows for an exact and independent variation of the renormalization and factorization scales.
For Tevatron and LHC we study its dependence 
on all scales, on the parton luminosity and on the top-quark mass 
using both the conventional pole mass definition as well as the running mass in the \msbar\ scheme.
We extract for the top-quark an \msbar\ mass of $m(\mu=m) = 160.0^{+3.3}_{-3.2}\GeV$, 
which corresponds to a pole mass of $\mt = 168.9^{+3.5}_{-3.4}\GeV$.
\vfill
\end{titlepage}

\newpage

The top-quark is the heaviest elementary particle discovered
so far and it is likely to be the most sensitive probe of the electroweak 
symmetry breaking. This is reflected in the fact that in many
extensions of the Standard Model the top-quark plays a special role.
The precise measurement of top-quark properties is thus an important
task for the Large Hadron Collider (LHC) (see e.g. Refs.~\cite{Bernreuther:2008ju,Incandela:2009pf}).  
One of the most basic quantities in that respect is the total cross
section which is currently measured at the Tevatron and will be
measured at the LHC. The precise measurements aimed for at the LHC are
asking for an equally precise theoretical prediction to compare with. 
In this Letter, we update and extend the predictions of Refs.~\cite{Moch:2008qy,Moch:2008ai}  
for the total hadronic cross section of top-quark pairs and its associated theoretical uncertainty. 
Related recent studies have also appeared in Refs.~\cite{Kidonakis:2008mu,Cacciari:2008zb,Nadolsky:2008zw}.
As a novel aspect of this Letter, we employ the \msbar\ definition for the top-quark mass and 
present the total cross section as a function of the running mass.
This allows the direct determination of an \msbar\ mass from Tevatron measurements 
for the total cross section~\cite{Abazov:2009ae}.

We start by recalling the relevant formulae for the total cross section of
top-quark hadro-production within perturbative Quantum Chromodynamics (QCD):
\begin{eqnarray}
  \label{eq:totalcrs}
  \sigma_{pp \to {t\bar t X}}(S,\mts) &=& 
  {\alpha_s^2 \over \mts}\, 
  \sum\limits_{i,j = q,{\bar{q}},g} \,\,\,
  \int\limits_{4\mts}^{S}\,
  ds \,\, L_{ij}(s, S, \mufs)\,\,
  f_{ij}(\rho,M,R)
  \, ,
\\
  \label{eq:partonlumi}
  L_{ij}(s,S,\mufs) &=& 
  {1\over S} \int\limits_s^S
  {d\shat\over \shat} 
  \phi_{i/p}\left({\shat \over S},\mufs\right) 
  \phi_{j/p}\left({s \over \shat},\mufs\right)
  \, ,
\end{eqnarray}
where $S$ and $\mt$ denote the hadronic center-of-mass energy squared and the
top-quark mass (here taken to be the pole mass), 
while $L_{ij}$ is the usual definition of the parton luminosity  
with the parton distributions (PDFs) $\phi_{i/p}$ at the factorization scale $\muf$.
The scaling functions $f_{ij}$ parameterize the hard partonic scattering process.
They depend only on dimensionless ratios of $\mt$, $\muf$, the renormalization scale $\mur$ and 
the partonic center-of-mass energy squared $s$, with the definitions $\rho = 4\mts/s$, 
$R = \murs/\mufs$ and $M = \mufs/\mts$. 
The radiative corrections to the scaling functions $f_{ij}$ at the next-to-leading order
(NLO)~\cite{Nason:1988xz,Beenakker:1989bq,Bernreuther:2004jv} are long known 
and recently even the complete analytic expressions have become available~\cite{Czakon:2008ii}.
In order to quantify the theory uncertainty also the next-to-next-to-leading order (NNLO) must be included. 
Presently, these NNLO corrections~\cite{Moch:2008qy,Moch:2008ai} are approximated  by the complete tower 
of the Sudakov logarithms and including all two-loop Coulomb corrections.

In the present study we include consistently the channels, $q{\bar q}$, $gg$ and $gq$ through NNLO 
and we provide parametrizations for all necessary scaling functions 
in the standard \msbar\ scheme for mass factorization. 
This allows for an easy handling in phenomenological applications.
Our phenomenological study reflects the latest measured value for the
top-quark mass~\cite{fnal:2009ec}, $\mt = 173.1^{+1.3}_{-1.3} \GeV$,  
and employs new PDF sets~\cite{Nadolsky:2008zw,Martin:2009iq}.
Let us briefly summarize the key aspects of our update with respect to
Refs. \cite{Moch:2008qy,Moch:2008ai}:
\begin{enumerate}
\item We use exact dependence on the renormalization and factorization scale.
  This allows for an independent variation of $\mur$ and $\muf$ (extending Ref.~\cite{Kidonakis:2001nj})  
  and is commonly considered as more reliable to establish the theoretical
  uncertainty of perturbative predictions 
  (see e.g. Ref.~\cite{Cacciari:2003fi}).
\item We perform the singlet-octet color decomposition consistently when matching our threshold expansion at NLO 
  using results of Refs.~\cite{Petrelli:1997ge,Hagiwara:2008df,Kiyo:2008bv}.
  The numerical impact on phenomenology at LHC and Tevatron turns out
  to be negligible, though.
\item We discuss those residual systematical uncertainties of our predictions for LHC and Tevatron, 
  which are inherent in the approach based on threshold resummation and we 
  comment on the size of unknown corrections.
\item We quantify the numerical impact of other known effects on the total cross section, such as 
  the electro-weak radiative corrections at NLO~\cite{Beenakker:1993yr,Bernreuther:2006vg,Kuhn:2006vh}
  and bound state corrections in QCD at threshold~\cite{Hagiwara:2008df,Kiyo:2008bv}.
\item We study the dependence of the total cross section on the definition of the mass parameter.
  For the conversion of the conventionally used pole mass $\mt$ (see Eq.~(\ref{eq:totalcrs})) 
  and the scale dependent \msbar\ mass $\mmu$ we exploit well-known relations to NNLO~\cite{Gray:1990yh} 
  (see also Refs.~\cite{Chetyrkin:1999qi,Melnikov:2000qh}).
  We investigate the apparent convergence of both definitions, $\mt$ and
  $\mmu$, in perturbation theory through NNLO.
\end{enumerate}
We also employ the analytic results for the NLO scaling functions~\cite{Czakon:2008ii}.  
As a net effect these lead to small improvements in the $gq$- and the $gg$-channel contributions of our NNLO prediction.

\subsection*{Theoretical setup}
The perturbative expansion of the scaling functions $f_{ij}$ in the strong coupling $\alpha_s$ 
up to two loops around $M = R = 1$, i.e. $\mt = \mur = \muf$, reads
\begin{eqnarray}
  \label{eq:scaling-fct}
  f_{ij}(\rho, 1, 1) &=& 
  f_{ij}^{(0)}(\rho) 
  + 4\pi \alpha_s 
    f_{ij}^{(10)}(\rho) 
  + \bigl( 4\pi \alpha_s\bigr)^2 
    f_{ij}^{(20)}(\rho) 
  \, ,
\end{eqnarray}
and the functions $f_{ij}^{(0)}$, $f_{ij}^{(10)}$ and $f_{ij}^{(20)}$ contain, 
at each order in $\alpha_s$, genuinely new information to be calculated from first principle in perturbation theory.
At Born level, 
\begin{eqnarray}
\label{eq:fqq0}
\fqqn &=& 
{\pi\, \beta\, \rho \over 27} [2 + \rho]
\, ,
\\
\label{eq:fgq0}
\fgqn &=& 
0
\, ,
\\
\label{eq:fgg0}
\fggn &=& 
{\pi\, \beta\, \rho \over 192} \biggl[
{1 \over \beta} (\rho^2 + 16 \rho + 16)  \ln\biggl(\frac{1+\beta}{1-\beta}\biggr)
-28 - 31 \rho
\biggr]
\, ,
\end{eqnarray}
where $\beta$ is the heavy quark velocity with $\beta=\sqrt{1-\rho}$. 
At NLO the known functions $f_{ij}^{(10)}$ can be described through parametrizations 
which are accurate at the per mille level.
Our one-loop fits use the following ansatz,
\begin{eqnarray}
\label{eq:fqq10}
\fqqon &=& 
   \frac{\rho \* \beta}{36\pi} \biggl[
     \frac{32}{3} \ln^2 \beta
   + \biggl(32 \ln 2 - \frac{82}{3}\biggr)\, \ln \beta
   - \frac{1}{12}\*\frac{\pi^2}{\beta}
   \biggr]
   + \beta  \rho a_0^{qq} + h(\beta,a_1,\ldots,a_{17}) 
\\
& &
   + \frac{1}{8\pi^2}(n_f-4) \fqqn \biggl[
     \frac{4}{3}  \ln 2 - \frac{2}{3} \ln \rho
   - \frac{10}{9}
   \biggr]
\ ,
\nonumber\\
\label{eq:fgq10}
\fgqon &=& 
   \frac{1}{16\pi}\beta^3\biggl[
   \frac{5}{9} \ln \beta
   + \frac{5}{6}\ln 2 - \frac{73}{108} 
    \biggr] 
   + h_{gq}^{(a)}(\beta,a_1,\ldots,a_{15})
\ ,
\\
\label{eq:fgg10}
\fggon &=& 
\frac{7\beta}{768\pi} \biggl[
     24 \ln^2 \beta
   +\biggl(72 \ln 2 - \frac{366}{7} \biggr)\ln \beta
   + \frac{11}{84}\*\frac{\pi^2}{\beta}
   \biggr] 
   + \beta  a_0^{gg} + h(\beta,a_1,\ldots,a_{17})
\\ & &
   + (n_f-4)\frac{\rho^2}{1024\pi} \biggl[
      \ln\biggl(\frac{1+\beta}{1-\beta}\biggr)
   - 2 \beta 
   \biggr]
\, ,
\nonumber
\end{eqnarray}
where $n_f$ denotes the number of light quarks and we have kept the complete dependence 
on $n_f$ in all parametrizations manifest.
The Sudakov logarithms $\ln \beta$ at threshold and the Coulomb corrections ($\sim 1/\beta$) 
in Eqs.~(\ref{eq:fqq10})--(\ref{eq:fgg10}) are exact~\cite{Nason:1988xz}. 
The constants $a_0^{ij}$ read
\begin{eqnarray}
\label{eq:a0qq}
  a_0^{qq} &=& \frac{299}{324\*\pi}-\frac{43}{1296}\*\pi 
  - \frac{121}{108}\*\frac{ \ln 2}{\pi}+\frac{16}{27}\*\frac{ \ln^2 2}{\pi}
  \approx 0.03294734
\, ,
\\
\label{eq:a0gg}
  a_0^{gg} &=& \frac{1111}{2304\*\pi}-\frac{283}{18432}\*\pi 
  -\frac{89}{128}\*\frac{ \ln 2}{\pi} + \frac{7}{16}\*\frac{ \ln^2 2}{\pi}
  \approx 0.01875287
\, .
\end{eqnarray}
They are known from the computation of the NLO QCD corrections to
hadro-production of quarkonium~\cite{Petrelli:1997ge} (see also Refs.~\cite{Hagiwara:2008df,Kiyo:2008bv}), 
where also details of the decomposition of $a_0^{gg}$ 
for color-singlet and color-octet states can be found.
The constants $a_0^{ij}$ in Eqs.~(\ref{eq:a0qq}) and (\ref{eq:a0gg}) 
emerge from Refs.~\cite{Petrelli:1997ge,Hagiwara:2008df,Kiyo:2008bv} 
by means of a simple Mellin transformation and agree with the values quoted in Ref.~\cite{Czakon:2008ii}.
The coefficients of the functions $h(\beta,a_1,\ldots,a_{17})$ and $h_{gq}^{(a)}(\beta,a_1,\ldots,a_{15})$ 
in Eqs.~(\ref{eq:fqq10})--(\ref{eq:fgg10}) 
are determined in a fit to the analytic expressions of Ref.~\cite{Czakon:2008ii}. 
Near threshold we have $h(\beta) = {\cal O}(\beta^2)$. 
More details are given in Eqs.~(\ref{eq:tophgg}) and (\ref{eq:tophgqa}).
Due to the larger number of parameters in the fit functions, 
it is evident that Eqs.~(\ref{eq:fqq10})--(\ref{eq:fgg10}) supersede earlier 
parametrizations~\cite{Nason:1988xz} with respect to accuracy.

At two-loop level we know the complete tower of Sudakov logarithms, $\ln^k \beta$ with $k=1,...,4$,
for the functions $\fqqtn$ and $\fggtn$ and, in addition, also the complete
Coulomb contributions, $\sim 1/\beta^2, 1/\beta$. 
The channel $gq$ is power suppressed near threshold relative to $q{\bar q}$ and $gg$.
However, extending soft gluon resummation to power suppressed quantities 
(see e.g. Refs.~\cite{Laenen:2008ux,Moch:2009mu}) and using Eq.~(\ref{eq:fgq10}), 
we can determine (at least) the leading term $\sim \ln^3 \beta$ of the function $\fgqtn$. 
We find
\begin{eqnarray}
\label{eq:fqq20}
\fqqtn &=& {\fqqn \over (16 \pi^2)^2} \*
  \biggl[
         {8192 \over 9} \* \ln^4 \beta
       + \biggl(  - {15872 \over 3} + {16384 \over 3} \* \ln 2 
       + {1024 \over 27} \* n_f \biggr) \* \ln^3 \beta
\\ &&
       + \biggl( 1046.4831 - 90.838135 \* n_f - 140.36771 \* {1 \over \beta} \biggr) \* \ln^2 \beta
\nonumber\\ &&
       + \biggl( 
       1029.8687 - 2.8903919 n_f 
       - 2 D^{(2)}_{Q{\bar Q}}
       + ( 54.038454 - 4.3864908 \* n_f ) \* {1 \over \beta} \biggr) \* \ln \beta
\nonumber\\ &&
       + 3.6077441 \* {1 \over \beta^2} 
       + ( 7.3996963 + 0.61492528 \* n_f ) \* {1 \over \beta} 
       + C^{(2)}_{q{\bar q}}
    \biggr]
\, ,
\nonumber\\
\label{eq:fgq20}
\fgqtn &=& {\beta^3 \over (16 \pi^2)^2} \* {65 \*\pi \over 54} \* \ln^3 (8 \* \beta^2)
\, ,
\\
\label{eq:fgg20}
\fggtn &=& {\fggn \over (16 \pi^2)^2} \*
  \biggl[ 
         4608 \* \ln^4 \beta
       + \biggl(  - {150400 \over 7} + 27648 \* \ln 2
       + {256 \over 3} \* n_f \biggr) \* \ln^3 \beta
\\ &&
       + \biggl( - 315.57218 
       - 119.35529 \* n_f 
       + 496.30011 \* {1 \over \beta} \biggr) \* \ln^2 \beta
\nonumber \\ &&
       + \biggl( 
       3249.2403 - 19.935233 n_f - 1.4285714 D^{(2)}_{Q{\bar Q}}
       + ( 286.67132 + 6.8930570 \* n_f ) \* {1 \over \beta} \biggr) \* \ln \beta
\nonumber \\ &&
       + 68.547138 \* {1 \over \beta^2} 
       - ( 192.10086 + 0.96631115 \* n_f ) \* {1 \over \beta} 
       + C^{(2)}_{gg}
    \biggr]
\, .
\nonumber 
\end{eqnarray}
Eqs.~(\ref{eq:fqq20}) and (\ref{eq:fgg20}) are exact up to the unknown constant terms 
$C^{(2)}_{q{\bar q}}$ and $C^{(2)}_{gg}$ of order ${\cal O}(\beta^0)$, 
whereas Eq.~(\ref{eq:fgq20}) receives further corrections of order ${\cal O}(\beta^3 \ln^2 \beta)$. 
Please note that the numerical coefficients in Eqs.~(\ref{eq:fqq20}) and (\ref{eq:fgg20}) 
have slightly changed compared to Ref.~\cite{Moch:2008qy}.
The $\ln^2 \beta$-terms in Eqs.~(\ref{eq:fqq20}) and (\ref{eq:fgg20}) are affected 
by using the exact coefficients~(\ref{eq:a0qq}) and (\ref{eq:a0gg}) in the matching at NLO.
The linear terms proportional to $\ln \beta$ in Eqs.~(\ref{eq:fqq20}) and (\ref{eq:fgg20}) 
contain genuine two-loop contributions.
Among those is the soft anomalous dimension $D^{(2)}_{Q{\bar Q}}$ (see Ref.~\cite{Moch:2008qy}).
Inserting the respective numerical coefficients we find $730.73916 + 23.776275 n_f$ 
in Eq.~(\ref{eq:fqq20}) and $3035.5764 - 0.88761378 n_f$ in Eq.~(\ref{eq:fgg20}). 
The latter value has changed with respect to Ref.~\cite{Moch:2008qy} 
due to a consistent separation of the color-singlet and color-octet contributions in $a_0^{gg}$ 
at NLO~\cite{Petrelli:1997ge,Hagiwara:2008df,Kiyo:2008bv}.
However the phenomenology is rather insensitive to this change and is only affected at the per mille level.
The Coulomb terms ($\sim 1/\beta$) in Eqs.~(\ref{eq:fqq20}) and (\ref{eq:fgg20}) 
contain all contributions from the two-loop virtual corrections.
Eq.~(\ref{eq:fgq20}) gives the leading (though formally power suppressed) contribution at two loops to the $gq$-channel.
We include $\fgqtn$ in our analysis for three reasons. 
Firstly, under evolution of the factorization scale the $gq$-channel mixes
with the two other channels and for a consistent study of the factorization scale dependence 
this channel also needs to be taken into account.
Next, the luminosity $L_{gq}$ in particular at LHC is sizeable and Eq.~(\ref{eq:fgq20}) offers 
a way to control its numerical impact at higher orders. 
Finally, Eq.~(\ref{eq:fgq20}) provides a first step towards a general study 
of power suppressed but logarithmically enhanced terms near threshold for
top-quark production.

In Eq.~(\ref{eq:totalcrs}) the dependence of the scaling functions $f_{ij}$ 
on the renormalization and factorization scales, $\mur$ and $\muf$, can also be made explicit.
Starting from the expansion in $\alpha_s$ through NNLO around $R=1$, i.e. $\mur = \muf$, 
we introduce 
\begin{eqnarray}
  \label{eq:scaling-fct-M}
  f_{ij}(\rho, M, 1) &=& 
  f_{ij}^{(0)}(\rho) 
  + 4\pi \alpha_s 
  \left\{ 
    f_{ij}^{(10)}(\rho) + L_M f_{ij}^{(11)}(\rho)
  \right\}
  \nonumber\\
  & &
  + \bigl( 4\pi \alpha_s\bigr)^2 
  \left\{ 
    f_{ij}^{(20)}(\rho) + L_M f_{ij}^{(21)}(\rho) + L_M^2 f_{ij}^{(22)}(\rho)
  \right\}
  \, ,
\end{eqnarray}
where we abbreviate $L_M = \ln(\mufs / \mts)$. 
The logarithmic tower in $M = \mufs/\mts$, 
that is all terms proportional to $L_M$ in Eq.~(\ref{eq:scaling-fct-M}) 
can be derived by renormalization group methods in a straight forward manner. 
The explicit results in the \msbar\ scheme for all channels read at NLO 
\begin{eqnarray}
\label{eq:fqq11}
  f_{ij}^{(11)} &=& 
  - \bigl(2 \* P_{ij}^{(0)} - 2 \* \b0 \* \one \bigr) \otimes   f_{ij}^{(0)}
  \, ,
  \\
\label{eq:fgq11}
  \fgqoo &=& 
  - \Pgq0 \otimes \fggn - {1 \over 2 \* n_f} \* \Pqg0 \otimes \fqqn
  \, ,
\end{eqnarray}
where $ij = \{q{\bar q}, gg\}$ in Eq.~(\ref{eq:fqq11}). At NNLO we have 
\begin{eqnarray}
\label{eq:fqq21}
  \fqqto &=& 
  - \bigl(2 \* \Pqq1 - 2 \* \b1\*\one\bigr) \otimes \fqqn 
  - \bigr(2 \* \Pqq0 - 3 \* \b0\*\one) \otimes \fqqon 
  - 2 \* \Pgq0 \otimes \fgqon 
  \, ,
  \\
\label{eq:fqq22}
  \fqqtt &=& 
  \biggl(2 \* \Pqq0\otimes\Pqq0 - 5 \* \Pqq0 \* \b0 + {1 \over 2 \* n_f} \* \Pqg0 \otimes \Pgq0 
  + 3 \* \b0^2 \*\one \biggr) \otimes \fqqn + \Pgq0\otimes \Pgq0 \otimes \fggn
  \, ,
  \\
\label{eq:fgq21}
  \fgqto &=& 
  - {1 \over 2 \* n_f} \* \Pqg0 \otimes \fqqon 
  - \bigl( \Pqq0 + \Pgg0 - 3 \* \b0\*\one \bigr) \otimes \fgqon 
  - \Pgq1 \otimes \fggn 
  - {1 \over 2 \* n_f} \* \Pqg1 \otimes \fqqn 
\\
&&\nonumber
  - \Pgq0 \otimes \fggon
  \, ,
  \\
\label{eq:fgq22}
  \fgqtt &=& 
  {1 \over 4 \* n_f} \* \Pqg0 \otimes \bigl( 3 \* \Pqq0 + \Pgg0  - 5 \* \b0 \*\one\bigr) \otimes \fqqn
  +   {1 \over 2} \* \Pgq0 \otimes \bigl( \Pqq0 + 3 \* \Pgg0 - 5 \* \b0\*\one \bigr) \otimes \fggn
  \, ,
  \\
\label{eq:fgg21}
  \fggto &=& 
  - \bigl(2 \* \Pgg1 - 2 \* \b1\*\one) \otimes \fggn 
  - \bigl(2 \* \Pgg0 - 3 \* \b0\*\one) \otimes \fggon 
  - 2 \* \Pqg0 \otimes \fgqon 
  \, ,
  \\
\label{eq:fgg22}
  \fggtt &=&   {1 \over 2 \* n_f} \* \Pqg0\otimes\Pqg0 \otimes \fqqn 
  + \bigl( \Pqg0 \otimes \Pgq0 + 2 \* \Pgg0\otimes\Pgg0 - 5 \* \Pgg0 \* \b0 + 3 \* \b0^2\*\one) \otimes \fggn
  \, .
\end{eqnarray}
In Eqs.~(\ref{eq:fqq11})--(\ref{eq:fgg22}) 
the $\otimes$ products have to be understood as standard convolutions and a sum 
over all active quarks and anti-quarks is implied as well.
The coefficients of the QCD $\beta$-function are given by 
\begin{equation}
  \label{eq:defbeta}
  \b0 = {1\over16 \pi^2} \left(11-{ 2\over 3} n_f\right)\, ,
  \qquad\qquad 
  \b1 = {1\over (16 \pi^2)^2 } \left(102-{38\over 3} n_f\right) 
  \, .
\end{equation}
The splitting functions $P_{ij}^{(l)}$ can be taken e.g. from Ref.~\cite{Moch:2004pa,Vogt:2004mw}.
At leading order they read 
\begin{eqnarray}
  (16 \pi^2)\* \Pqq0(x) &=&
  {4 \over 3} \* \bigg( \,
  {4 \over 1 - x} - 2 - 2 \* x + 3\* \delta(1 - x)
  \,\bigg)
  \, ,
\label{eq:Pqq0}
\\
  (16 \pi^2)\* \Pqg0(x) &=&
  2\, \* n_f \, \* (1 - 2x + 2x^{\,2})
  \, ,
\label{eq:Pqg0}
\\
  (16 \pi^2)\* \Pgq0(x) &=&
  {4 \over 3} \* \bigg( \, {4 \over x} - 4 + 2 \* x
  \,\bigg)
  \, ,
\label{eq:Pgq0}
\\
  (16 \pi^2)\* \Pgg0(x) &=& 
  3  \*  \bigg(
  {4 \over 1-x} + {4 \over x} - 8 + 4 \* x - 4 \* x^{\,2}
  + {11 \over 3} \* \delta(1 - x)
  \bigg)
  - {2 \over 3} \, \* n_f \, \* \delta(1 - x)
  \, ,
\label{eq:Pgg0}
\end{eqnarray}
where the factor $(16 \pi^2)$ accounts for the normalization used in Eqs.~(\ref{eq:fqq11})--(\ref{eq:fgg22}). 
In general, we have 
$P^{(l)}_{ij}({\footnotesize \mbox{this~article}}) = (16 \pi^2)^{-(l+1)}\, P^{(l)}_{ij}({\footnotesize \mbox{Refs.~\cite{Moch:2004pa,Vogt:2004mw}}})$.
Please also note the explicit factor of $(2 \* n_f)^{-1}$ in Eqs.~(\ref{eq:fgq11})--(\ref{eq:fgg22}), 
which is due to the definition of $\Pqg0$ in Eq.~(\ref{eq:Pqg0}) and $\Pqg1$ in Ref.~\cite{Vogt:2004mw}.
Simple fully analytic expressions for $\fqqoo$, $\fgqoo$ and $\fggoo$ are long known~\cite{Nason:1988xz}
and precise fits for all scaling functions $f_{ij}^{(21)}$, $f_{ij}^{(22)}$ in Eqs.~(\ref{eq:fqq21})--(\ref{eq:fgg22}), 
typically to per mille accuracy, are presented in the Appendix in
Eqs.~(\ref{eq:topf21qq})--(\ref{eq:topf22gg}) and Tabs.~\ref{tab:qqfit}--\ref{tab:ggfit}.
Finally, the complete scale dependence for $f_{ij}(\rho, M, R)$ in Eq.~(\ref{eq:totalcrs}) 
with $\mur \neq \muf$ is easily obtained as
\begin{eqnarray}
  \label{eq:scaling-Rqq}
  f_{ij}(\rho, M, R) &=& 
  f_{ij}(\rho, M, 1) 
  + 4\pi \alpha_s \left\{ 2 \* \b0 \* L_R \* f_{ij}^{(0)} \right\} 
  \\
  & &
  + \bigl( 4\pi \alpha_s\bigr)^2 
  \left\{ 
       3 \* \b0 \* L_R \* f_{ij}^{(10)}
     + 2 \* \b1 \* L_R \* f_{ij}^{(0)}
     + 3 \* \b0 \* L_R \*L_M \* f_{ij}^{(11)}
     + 3 \* \b0^2 \* L_R^2 \* f_{ij}^{(0)}   
  \right\}
  \, ,
  \nonumber\\
  \label{eq:scaling-Rgq}
  f_{gq}(\rho, M, R) &=& 
  f_{gq}(\rho, M, 1) 
  + \bigl( 4\pi \alpha_s\bigr)^2 
  \left\{ 
      3 \* \b0 \* L_R \* \fgqon
    + 3 \* \b0 \* L_M \* \fgqoo
  \right\}
  \, ,
%
%
\end{eqnarray}
where $L_R = \ln(\murs / \mufs)$ and $ij = \{q{\bar q}, gg\}$ in Eq.~(\ref{eq:scaling-Rqq}).

\subsection*{Phenomenological applications}
We are now in the position to address the phenomenological consequences. 
The approximate NNLO prediction which includes exact dependence on all scales 
is based on Eqs.~(\ref{eq:fqq20})--(\ref{eq:fgg20}), 
(\ref{eq:fqq21})--(\ref{eq:fgg22}) and (\ref{eq:scaling-Rqq}), (\ref{eq:scaling-Rgq}).
If not otherwise stated, the top-quark mass is the pole mass at $\mt = 173\GeV$.

\begin{figure}[ht!]
\centering
\vspace*{10mm}
\subfigure[\label{subfig:totalXsection:left}]
    {
      \scalebox{0.9}{\includegraphics[%
          bbllx=77pt,bblly=541pt,bburx=298pt,bbury=767pt]{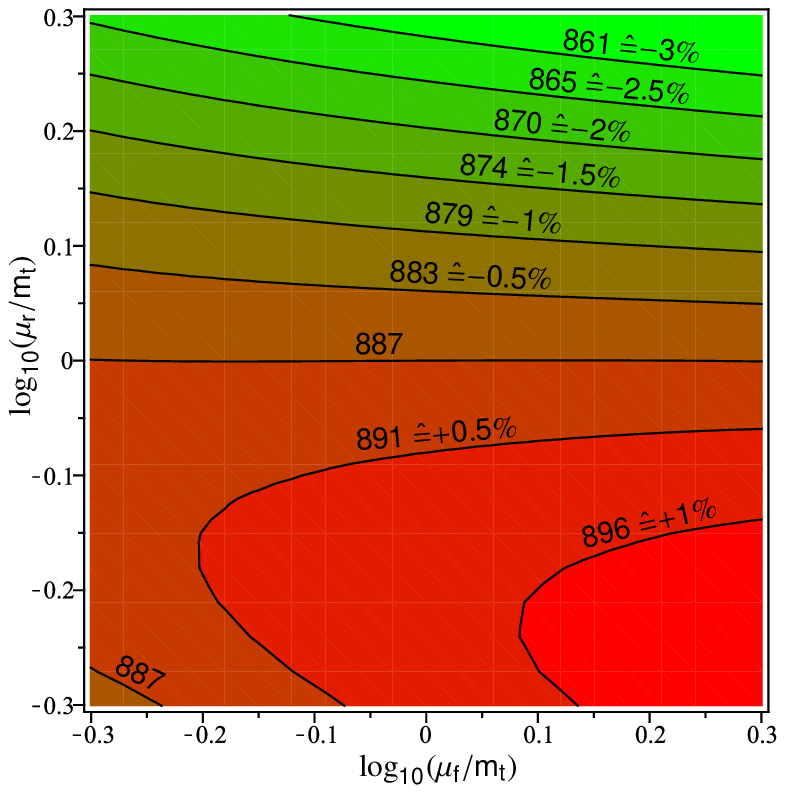}}
    }
\hspace*{10mm}
\subfigure[\label{subfig:totalXsection:right}]
    {
      \scalebox{0.9}{\includegraphics[%
          bbllx=77pt,bblly=541pt,bburx=298pt,bbury=767pt]{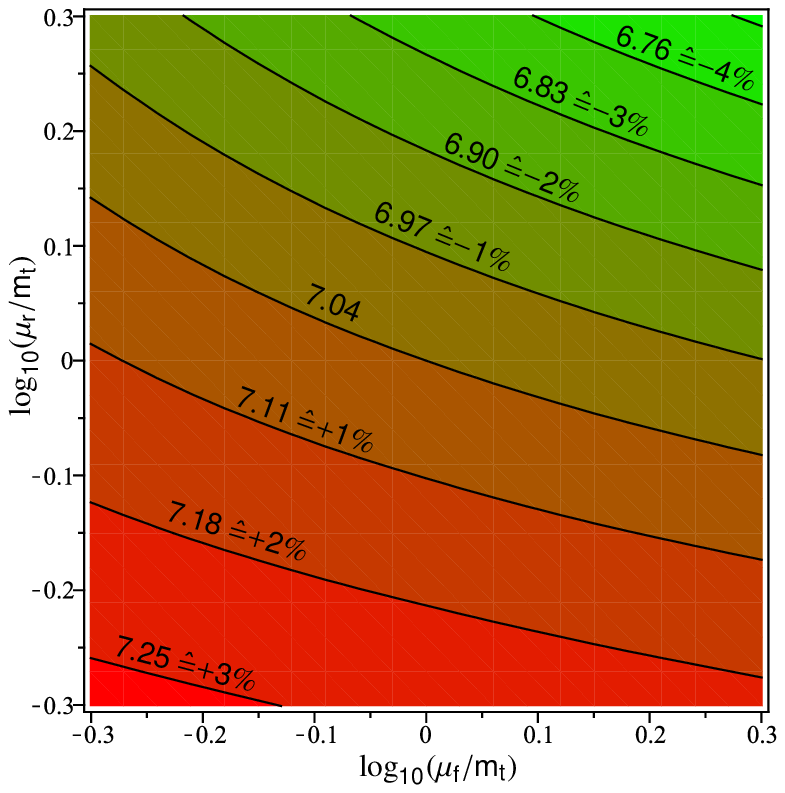}}
    }
\caption{\small
  \label{fig:totalXsection}
  The contour lines of the total hadronic cross section 
  from the independent variation of renormalization and factorization scale 
  $\mur$ and $\muf$ for LHC with $\sqrt{S} = 14\TeV$ (left) 
  and Tevatron with $\sqrt{S} = 1.96\TeV$ (right) with
  MSTW 2008~\cite{Martin:2009iq}.
  The cross sections are normalized to the values at $\mur = \muf = \mt$ 
  and the range corresponds to $\mur,\muf \in [\mt/2,2\mt]$.
}
\end{figure}
Let us start with the scale dependence of the NNLO cross section as shown in Fig.~\ref{fig:totalXsection}.
Our study of the theoretical uncertainty allows us to assess the effect of independent variations 
of the renormalization and factorization scale $\mur$ and $\muf$ 
in the scaling functions for the hard partonic scattering process in Eq.~(\ref{eq:totalcrs}).
In doing so, one should keep in mind however, 
that all currently available PDF sets from global fits always fix $\mur = \muf$. 
We define the theory uncertainty arising from the independent variation 
of $\mur$ and $\muf$ in the standard range $\mur,\muf \in [\mt/2,2\mt]$ as 
\begin{eqnarray}
  \label{eq:range}
  \mbox{min}~ \sigma(\mur,\muf)
  \, \le \,
  \sigma(\mt)
  \, \le \,
  \mbox{max}~\sigma(\mur,\muf)
  \, .
\end{eqnarray}
The contour lines of the total cross section for the LHC and Tevatron arising
from this procedure are shown in Fig.~\ref{fig:totalXsection}. 
The standard range $\mur,\muf \in [\mt/2,2\mt]$ corresponds to the region displayed 
in Fig.~\ref{fig:totalXsection} because of $\log_{10}(2) \approx 0.3$. 
We have normalized all results to the value at $\mur = \muf = \mt$ 
and the variation with fixed scales $\mur = \muf$ proceeds 
along the diagonal from the lower left to upper right in the plots. 
For Tevatron in Fig.~\ref{subfig:totalXsection:right} we see, 
that the gradient in the complete $(\mur,\muf)$-plane is almost parallel to 
this diagonal, thus the uncertainty according to Eq.~(\ref{eq:range}) remains 
$ -5\% \le \Delta \sigma \le +3\%$ 
in very good agreement with previous results~\cite{Moch:2008qy,Moch:2008ai}.
At LHC in Fig.~\ref{subfig:totalXsection:left} the maximal deviations 
in the $(\mur,\muf)$-plane are $-3\%$ located at $(2\mt,2\mt)$ 
and $+1\%$ at roughly $(\mt/2,2\mt)$, so that the uncertainty range~(\ref{eq:range}) 
becomes slightly larger, $ -3\% \le \Delta \sigma \le +1\%$ 
compared to what has been derived before with a fixed scale ratio $\mur = \muf$.
Very similar numbers for both colliders are obtained with the PDF set CTEQ6.6~\cite{Nadolsky:2008zw}.
Also recall, that we include the $gq$-channel through two loops. 
Thanks to Eqs.~(\ref{eq:fqq21})--(\ref{eq:fgg22}) we control the exact scale dependence 
also for all contributions proportional to the parton luminosity $L_{gq}$.
We conclude from Fig.~\ref{fig:totalXsection} 
that the theoretical uncertainty due to $\mur$ and $\muf$ variation 
is well estimated by case of identical scales $\mur = \muf$.

\begin{figure}[ht!]
\centering
\vspace*{10mm}
\subfigure[\label{subfig:totalerror:left}]
    {
      \scalebox{0.32}{\includegraphics{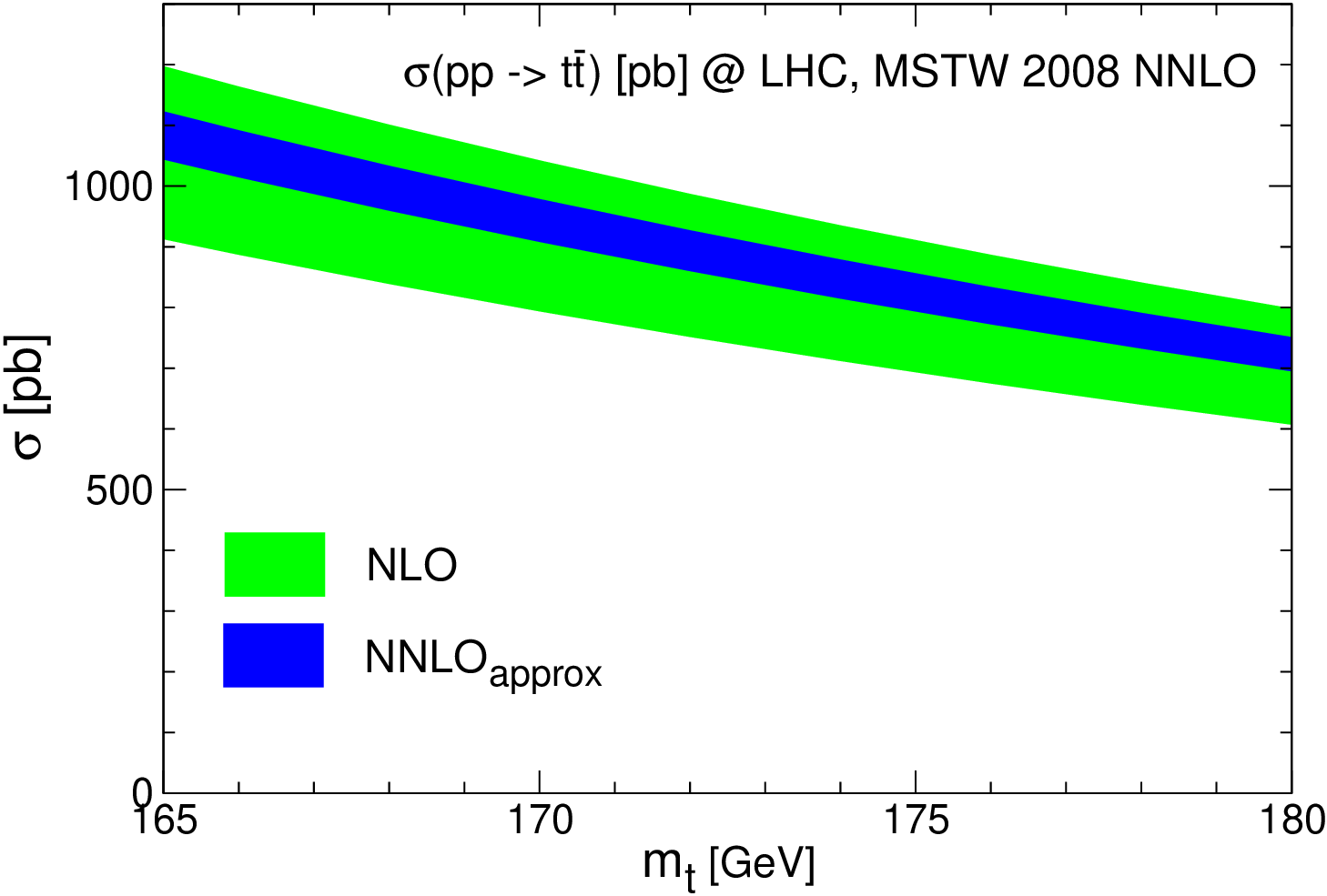}}
    }
\hspace*{2mm}
\subfigure[\label{subfig:totalerror:right}]
    {
      \scalebox{0.32}{\includegraphics{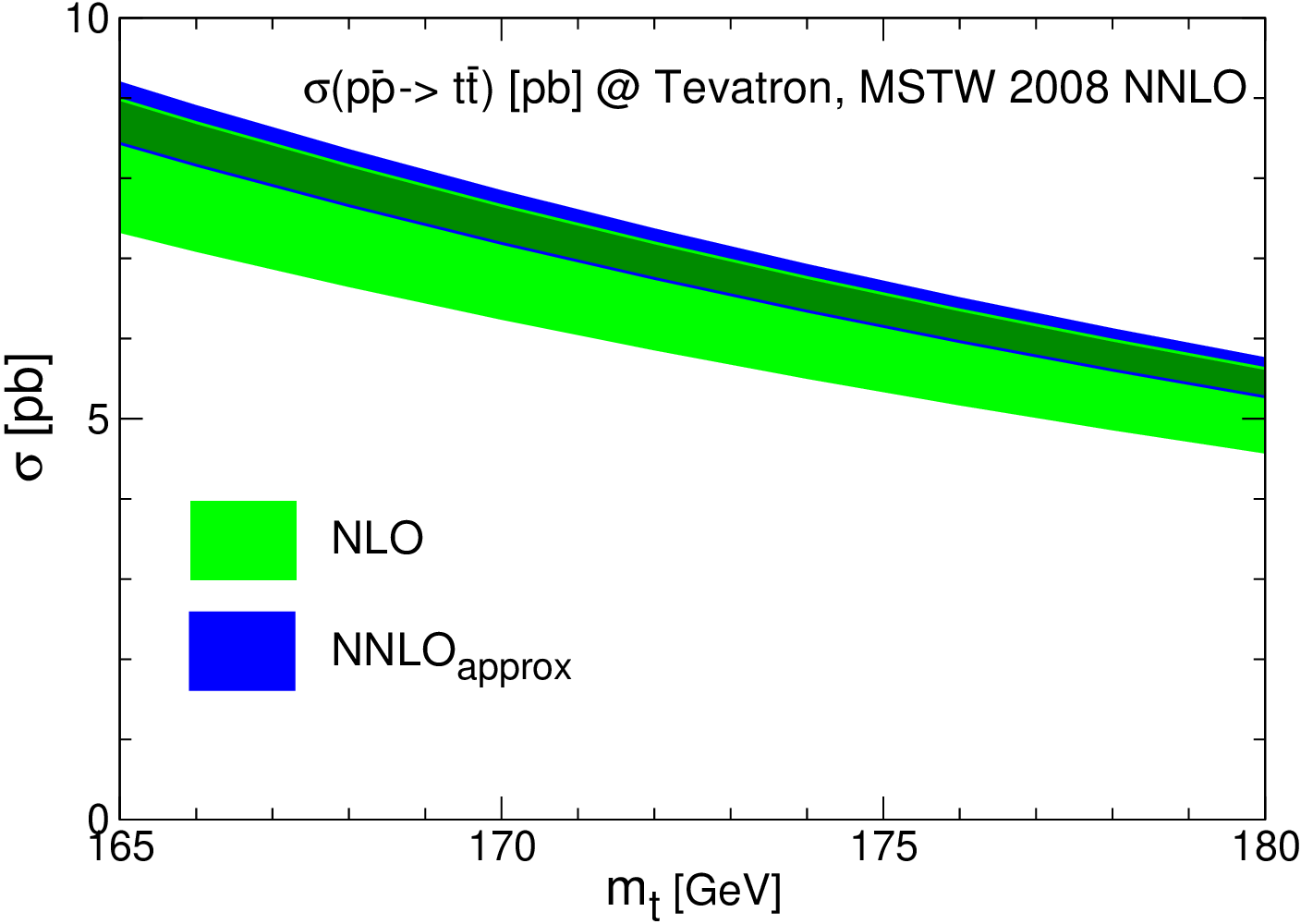}}
    }
\caption{\small
  \label{fig:totalerror}
  The mass dependence of the total cross section at NLO (green) and
  approximate NNLO (blue) order for LHC at $\sqrt{S} = 14\TeV$ (left) and Tevatron at $\sqrt{S} = 1.96\TeV$ (right) 
  and the PDF set MSTW 2008~\cite{Martin:2009iq}.
  The bands denote the theoretical uncertainty from scale variation keeping
  $\mur = \muf$ and the
  PDF uncertainty in the range $[\mt/2,2\mt]$.
In Fig.~\ref{subfig:totalerror:right} for the Tevatron 
the NLO and approximate NNLO bands overlap only partially giving rise to the 
corresponding medium band (dark green).
}
\end{figure}
In order to quantify the PDF uncertainty we apply the standard definition 
\begin{equation}
  \label{eq:pdferr}
  \Delta \sigma =
  \frac{1}{2} \, \sqrt{\sum_{k=1,n_{PDF}} \, (\sigma_{k+} - \sigma_{k-})^2}
  \, ,
\end{equation}
which determines $\Delta \sigma$ from the variations $\sigma_{k\pm}$ 
of the cross section with respect to the $k$ parameters of the PDF fit.
Typically the PDF error is added linearly to the theory uncertainty obtained from 
the scale variation. This is the commonly adopted choice and we employ 
the PDF sets CTEQ6.6~\cite{Nadolsky:2008zw} and MSTW 2008~\cite{Martin:2009iq}. 
The latter set gives two uncertainties at different confidence levels (CL), one at 68\%~CL and the second at 90\%~CL.
Throughout this study we use 68\%~CL only.
Moreover in the chosen interval $\mu \in [\mt/2,2\mt]$ for a given $\mu = \mur = \muf$ 
the error $\Delta \sigma(\mu)$ in Eq.~(\ref{eq:pdferr}) has only a very weak scale dependence. 
That is to say we find to good accuracy 
$\Delta \sigma(\mu=\mt/2) \simeq \Delta \sigma(\mu=\mt) \simeq \Delta \sigma(\mu=2\mt)$
so that a determination of $\Delta \sigma$ at the central scale $\mu=\mt$
should suffice for all practical purposes.

In Fig.~\ref{fig:totalerror} we show the mass dependence of the total cross section, 
comparing the NLO and our approximate NNLO prediction. 
The band summarizes the total theoretical uncertainty from the linear combination
of the scale uncertainty for the case $\mur = \muf$ 
and the PDF uncertainty Eq.~(\ref{eq:pdferr}).
We display the LHC and Tevatron predictions 
using the MSTW 2008 PDF set~\cite{Martin:2009iq}.
The improvement of the NNLO prediction is manifest for both colliders. 

\bigskip

Next, we discuss the sources of remaining systematical uncertainties.
Undoubtly, a complete calculation of the complete NNLO QCD corrections to
hadronic top-quark pair-production would be highly desirable 
(see Refs.~\cite{Czakon:2007wk,Czakon:2007ej,Korner:2008bn,Kniehl:2008fd,%
Anastasiou:2008vd,Bonciani:2008nd,Czakon:2008zk,Bonciani:2009nb} 
for progress in this direction).
This lacking, the main systematic uncertainty of our approximate NNLO result
are the sub-leading terms in the scaling function $f_{ij}^{(20)}$.
They might become accessible by extending the approach of Refs.~\cite{Laenen:2008ux,Moch:2009mu} 
or else could be modeled through power suppressed terms in Mellin space 
(see e.g. the scheme $A=2$ in Ref.~\cite{Bonciani:1998vc}).
By including the leading term for $\fgqtn$ we have taken a first step in this direction 
and we have found numerically small effects only.
In order to quantify our systematical uncertainty, we adopt the following prescription:
We compute the ratio $\sigma_{\rm NLL+Coul}/\sigma_{\rm exact}$ at one loop, 
where $\sigma_{\rm NLL+Coul}$ only contains the Sudakov logarithms and the Coulomb terms in $\fqqon$ and $\fggon$, 
i.e. the content of the square brackets in Eqs.~(\ref{eq:fqq10}) and (\ref{eq:fgg10}). 
This checks how well the exact hadronic cross section in Eq.~(\ref{eq:totalcrs}) is approximated, 
if only the threshold approximation enters in the convolution with the parton luminosities.
Typically we find $\sigma_{\rm NLL+Coul}/\sigma_{\rm exact} \gsim 0.7 (0.9)$ for the LHC (Tevatron).
If translated to the genuine two-loop contribution (see e.g. Tab.~\ref{tab:crsnumbers} below), 
then a systematic uncertainty of ${\cal O}(30\%)$ implies a cross section uncertainty 
of $\Delta \sigma \simeq {\cal O}(15)$~pb at LHC and of $\Delta \sigma \simeq {\cal O}(0.2)$~pb at Tevatron.
These numbers are corroborated by other observations, like the generally small
impact of the $gq$-channel which is entirely sub-leading.

\begin{figure}[ht!]
\centering
\vspace*{10mm}
\subfigure[\label{subfig:scaledep:left}]
    {
        \scalebox{0.32}{\includegraphics{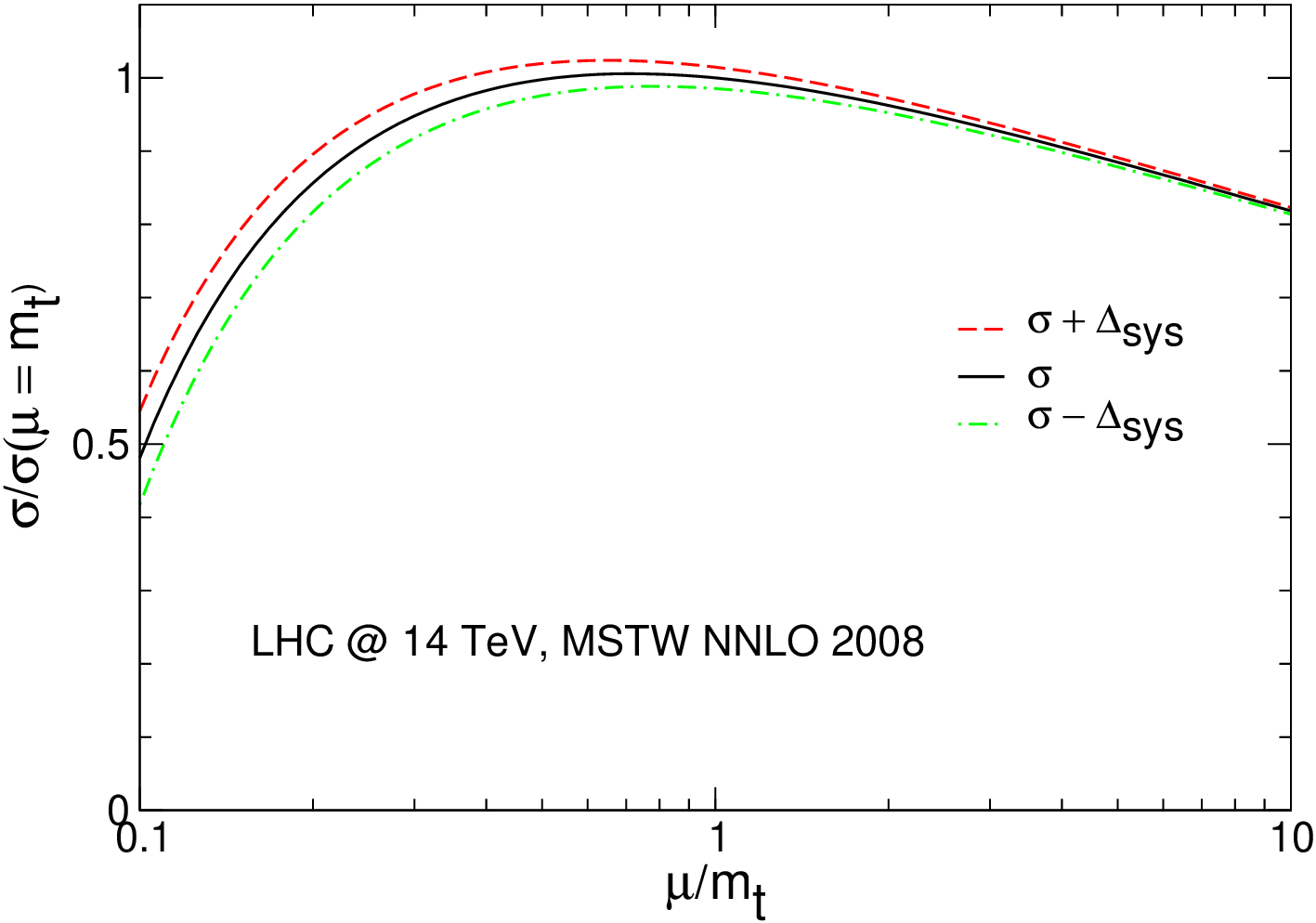}}
    }
\hspace*{3mm}
\subfigure[\label{subfig:scaledep:right}]
    {
        \scalebox{0.32}{\includegraphics{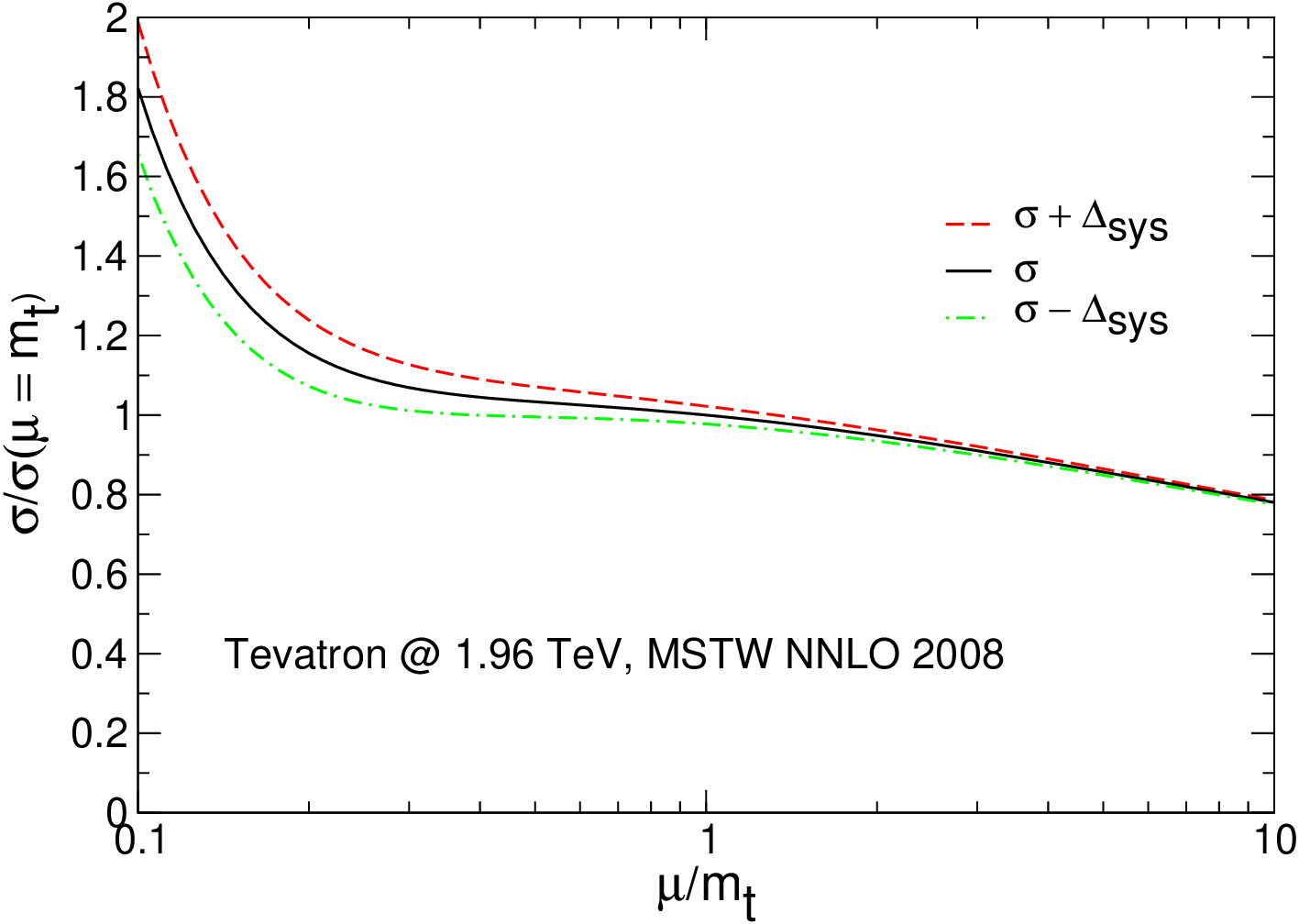}}
    }
\caption{\small
  \label{fig:scaledep}
 The scale dependence of the approximate NNLO cross section $\sigma_{\rm NNLO}$ 
 for the choice $\mu = \mur = \muf$ 
 using the MSTW 2008 PDF set~\cite{Martin:2009iq} at LHC with $\sqrt{S} = 14\TeV$ (left) 
 and Tevatron with $\sqrt{S} = 1.96\TeV$ (right).
 $\Delta_{\rm sys}$ denotes the estimated systematic uncertainty of our 
 threshold approximation at NNLO and all results are normalized to the value
 of $\sigma(\mu = \mt)$.
}
\end{figure}
How does this affect the previous discussion of the scale dependence? 
Let us define a systematic uncertainty $\Delta_{\rm sys}$ obtained from 
a variation of the scaling functions $f_{ij}^{(20)}$ 
in Eqs.~(\ref{eq:fqq20})--(\ref{eq:fgg20}) by $\pm 30 \%$.
All other scaling functions are known exactly anyway. 
The result for our NNLO cross section 
(always normalized to the value at $\mu = \mt$ and $\Delta_{\rm sys}=0$) 
is shown in Fig.~\ref{fig:scaledep} for $\mu = \mur = \muf$ and the MSTW 2008 PDF set~\cite{Martin:2009iq}.
It is obvious that the predictions are very stable within the standard range $\mu \in [\mt/2,2\mt]$ 
for all cases, i.e. $\Delta_{\rm sys}=0$ and $\sigma \pm \Delta_{\rm sys}$.
For the case $\Delta_{\rm sys}=0$ we find a variation of $-3\% \le \Delta \sigma \le +0.5\% $ for LHC 
and $ -5\% \le \Delta \sigma \le +3\%$ for Tevatron (compatible with Fig.~\ref{fig:totalXsection}) 
and similar numbers for the other two cases, $\sigma \pm \Delta_{\rm sys}$.

In the present analysis, we have also neglected the effect of the new parton channels 
$qq$, ${\bar q}\,{\bar q}$ and $q_i{\bar q}_j$ (for unlike flavors $i \neq j$), 
which come in through real emission at NNLO only. 
Important insight can be gained here from the recent calculation~\cite{Dittmaier:2007wz,Dittmaier:2008uj} 
of the radiative corrections for $t{\bar t}+$1-jet production at NLO, 
because they represent a significant part of the complete NNLO corrections for inclusive top-quark pair-production. 
At NLO $t{\bar t}+$1-jet production contains the one-loop one-parton real emissions as well as the double real
emission processes, and the latter also include the above mentioned new channels. 
It was found that the radiative corrections at the scale $\mur = \muf=\mt$ are rather small. 
Depending on the kinematical cuts (e.g. on the transverse momentum of the jet)
they amount to ${\cal O}(20)$~pb at LHC and to ${\cal O}(0.2)$~pb at Tevatron 
(see Ref.~\cite{Dittmaier:2008uj}). 
This provides further evidence that the hard corrections to the inclusive top-quark pair-production
at NNLO are indeed not large and it supports the estimate of our systematical uncertainty.

\bigskip

To summarize, our approximate NNLO prediction leaves us with a rather small residual theoretical uncertainty 
based on the scale variation. 
It is also worth stressing that the numerical impact of our theory improvements 
in Eqs.~(\ref{eq:fqq20})--(\ref{eq:fgg20}) and (\ref{eq:fqq21})--(\ref{eq:fgg22}) is rather small, 
which again nicely illustrates the stability of the approximate NNLO predictions. 
\begin{table}[ht!]
\begin{center}
  \begin{tabular}{l|c|c}
      &\quad~LHC~$\quad$  
      &Tevatron
      \\ \hline 
      $\sigma_{\rm LO} [\pb]$
      & 583.7
      & 5.820
      \\
      $\sigma_{\rm NLO} [\pb]$
      & 877.4
      & 7.229
      \\
      $\sigma_{\rm NNLO}[\pb]$ {\footnotesize (this work)}
      & 923.0
      & 7.814
      \\
      $\sigma_{\rm NNLO}[\pb]$ {\footnotesize (Refs.~\cite{Moch:2008qy,Moch:2008ai})}
      & 920.5
      &  7.810
      \\
      \hline 
  \end{tabular}
\end{center}
\caption{\small
  \label{tab:crsnumbers}
 The  LO, NLO and approximate NNLO prediction for the 
 total cross section at LHC ($\sqrt{S} = 14\TeV$) and Tevatron ($\sqrt{S} = 1.96\TeV$) 
 using $\mt = 171\GeV$, the PDF set CTEQ6.6~\cite{Nadolsky:2008zw} and 
 $\mur = \muf = \mt$.
 For comparison we also give the previous numbers of 
 Refs.~\cite{Moch:2008qy,Moch:2008ai}.
}
\end{table}
In Tab.~\ref{tab:crsnumbers} we compare with our previous numbers~\cite{Moch:2008qy,Moch:2008ai} 
for the CTEQ6.6~\cite{Nadolsky:2008zw} set at $\mt = 171 \GeV$ and $\mur = \muf = \mt$.
At Tevatron, we find no changes, as the cross section is entirely dominated
by the $q{\bar q}$-channel at parton kinematics close to threshold. 
At LHC, there is a small net change of $2.5$~pb in the prediction. 
Here, the effect of the improved NLO matching (in particular the $\ln^2 \beta$
term in Eq.~(\ref{eq:fgg20}) depending on the constant $a_0^{gg}$ in Eq.~(\ref{eq:a0gg})) 
and the leading two-loop term $\fgqtn$ in Eq.~(\ref{eq:fgq20}) in the $gq$-channel 
partially compensate. 
The numerical impact of the exact color decomposition at NLO (affecting the $\ln \beta$
term in Eq.~(\ref{eq:fgg20})) is completely negligible.

As a central result of the current studies we quote our approximate NNLO prediction 
at LHC ($\sqrt{S} = 14\TeV$) and Tevatron ($\sqrt{S} = 1.96\TeV$) for a pole mass of $\mt = 173 \GeV$. 
For the MSTW 2008 set~\cite{Martin:2009iq} we have
\begin{eqnarray}
  \label{eq:lhcnumber}
\sigma_{\rm LHC} &=& 887\,\,\pb\quad^{+9}_{-33}
\,\,\,\,\,\, \pb\,\,({\rm scale})\quad^{+15}_{-15}
\,\,\,\,\,\, \pb\,\,(\rm{MSTW 2008})
\, ,
\\
  \label{eq:tevnumber}
\sigma_{\rm Tev} &=&  7.04\, \pb\quad^{+0.24}_{-0.36}
\, \pb\,\,({\rm scale})\quad^{+0.14}_{-0.14}
\, \pb\,\,(\rm{MSTW 2008})
\, ,
\end{eqnarray}
and for CTEQ6.6~\cite{Nadolsky:2008zw},
\begin{eqnarray}
  \label{eq:cteqlhcnumber}
\sigma_{\rm LHC} &=& 874\,\, \pb\quad^{+9}_{-33}
\,\,\,\,\,\, \pb\,\,({\rm scale})\quad^{+28}_{-28}
\,\,\,\,\,\, \pb\,\,(\rm{CTEQ6.6})
\, ,
\\
  \label{eq:cteqtevnumber}
\sigma_{\rm Tev} &=&  7.34\, \pb\quad^{+0.24}_{-0.38}
\, \pb\,\,({\rm scale})\quad^{+0.41}_{-0.41}
\, \pb\,\,(\rm{CTEQ6.6})
\, .
\end{eqnarray}
Please note, that the MSTW 2008 set~\cite{Martin:2009iq} is based on a global analysis to NNLO in QCD  
while CTEQ6.6~\cite{Nadolsky:2008zw} performs a fit to NLO only.
Therefore, the two PDF sets return slightly different default values for the coupling constant $\alpha_s$. 
While the LHC predictions of both sets are largely in agreement for these choices of $\alpha_s$, 
the difference in the Tevatron predictions can be attributed to differences in the
parametrization of the light quark PDFs at large $x$.
In addition, there is a systematical uncertainty in Eqs.~(\ref{eq:lhcnumber})--(\ref{eq:cteqtevnumber}) 
estimated to be ${\cal O}(2 \%)$ due to unknown NNLO contributions, i.e. the exact expression for $f_{ij}^{(20)}$ 
in Eqs.~(\ref{eq:fqq20})--(\ref{eq:fgg20}).
One could of course argue that the accuracy of a given PDF set should match the accuracy of the 
theoretical prediction of the partonic cross section. 
However, we would like to disentangle the shift orginating from corrections to the hard parton scattering 
(which is the main subject of our paper) from PDF effects. 
Therefore we always use the same order in perturbation theory as far as the
chosen PDFs are concerned 
(cf. Tab.~\ref{tab:crsnumbers} and also Appendix B of Ref.~\cite{Moch:2008qy}).

For applications, the mass dependence of the hadronic cross section~(\ref{eq:totalcrs}) 
is conveniently parameterized by the following simple fit formula
\begin{eqnarray}
  \label{eq:hadrofit}
  \sigma(\mt,\mu) = a + b x + c x^2 + d x^3 +e x^4 + f x^5 + g x^6
\, ,
\end{eqnarray}
with $\mu = \mur = \muf$, $x = (\mt/\GeV - 173)$ and the scale choices $\mu = \mt, 2 \mt, \mt/2$.
For the cross section at LHC ($\sqrt{S} = 14\TeV$ and $\sqrt{S} = 10\TeV$) and Tevatron ($\sqrt{S} = 1.96\TeV$) 
all fit coefficients are listed in Tabs.~\ref{tab:hadrofitcteq} and \ref{tab:hadrofitmstw}, where
we have used the PDF sets CTEQ6.6~\cite{Nadolsky:2008zw} 
and MSTW 2008~\cite{Martin:2009iq}.
In the mass range $150\GeV \leq \mt \leq 220 \GeV$, the accuracy of the fit is
always better than $2.5$\textperthousand.

Let us briefly mention also other types of radiative corrections, which have not been considered here, 
e.g. in Eqs.~(\ref{eq:lhcnumber})--(\ref{eq:hadrofit}).
Within QCD these are bound state effects for the $t {\bar t}$-pair near threshold~\cite{Hagiwara:2008df,Kiyo:2008bv}.
They affect the total cross section at LHC of the order ${\cal O}(10)$~pb 
and, even more so, differential distributions in the invariant mass of the top-quark pair. 
At Tevatron, due to the dominance of the $q{\bar q}$-channel in the color-octet
configuration, they are negligible though.
Precision analyses at the percent level naturally need to consider  
also the electro-weak radiative corrections at NLO~\cite{Beenakker:1993yr,Bernreuther:2006vg,Kuhn:2006vh}.
Depending on the Higgs mass they cause a decrease relative to the LO cross section 
between ${\cal O}(2 \%)$ for a light Higgs ($m_h=120$GeV) 
and ${\cal O}(2.5 \%)$ for a heavy Higgs ($m_h=1000$GeV) at the LHC.
This amounts to a negative contribution $\Delta \sigma_{\rm EW} \simeq {\cal O}(10-15)$~pb.
At the Tevatron, the electro-weak radiative corrections are almost zero for a light Higgs ($m_h=120$GeV) and 
give a negative contribution of order ${\cal O}(1 \%)$, 
i.e. $\Delta \sigma_{\rm EW} \simeq {\cal O}(0.05)$~pb for a heavy
Higgs ($m_h=1000$GeV).

\subsection*{The top-quark mass in the \msbar\ scheme}
So far we have used the pole mass of the top-quark as a definition of the mass parameter.
However, it is well-known that the concept of the pole mass has intrinsic theoretical limitations 
owing to the fact that the top-quark is a colored object. 
As such it does not appear as an asymptotic state of the $S$-matrix due to confinement. 
In other words the $S$-matrix does not have a pole in the top-quark channel. 
The impact of different mass renormalizations has been investigated in great detail 
in the context of top-quark mass measurements at a future linear collider 
where a precision of the order of a few hundred MeV is envisaged.
In particular it has been shown that indeed the conceptual limitations 
of the pole mass lead to a poorly behaved perturbative series.
A class of alternative mass definitions, so-called short distance masses, offer
a solution to this problem, e.g. the $1S$-mass or the potential subtracted (PS) mass (see e.g. Ref.~\cite{Hoang:2000yr}).

In the following we study the impact of the conversion from
the pole mass scheme to the \msbar\ scheme (see
Ref.~\cite{Gray:1990yh,Chetyrkin:1999qi,Melnikov:2000qh} and references therein) 
for the total cross section of top-quark hadro-production.
This is a novel feature and, in principle, the cross section in terms of the \msbar\ mass 
can be used for a direct measurement of the running mass at a high scale. 
This is similar to the case of $b$-quark production at LEP 
(see Refs.~\cite{Rodrigo:1997gy,Brandenburg:1999nb} and \cite{Barate:2000ab,Abbiendi:2001tw,Abdallah:2005cv,delphi:2008wy}). 
Let us first describe briefly how we translate the predictions for the total cross section 
from the pole mass to the \msbar\ mass scheme. 
The starting point is the well-known relation between the pole mass $\mt$ and the \msbar\ mass $\mmu$ to NNLO:
\begin{equation}
  \label{eq:mpole-mbar}
  \mt = \mmu \* \left(1 + a_s(\mur) d^{(1)} + a_s(\mur)^2 d^{(2)}\right)
  \, ,
\end{equation}
with $a_s = \alpha_s^{(n_f=5)}/\pi$ (i.e. five active flavors) and coefficients 
$d^{(i)}$, which in general depend on the ratio ${\mur^2/\mmu^2}$,
\begin{eqnarray}
  \label{eq:d1def}
  d^{(1)} &=& {4\over 3} + \lm
  \, ,
  \\
  \label{eq:d2def}
  d^{(2)} &=& {307\over 32} + 2 \* \z2 + {2\over 3} \* \z2 \* \ln 2 - {1\over 6}\*\z3 
  + {509\over 72}\*\lm
  + {47\over 24}\*\lm^2 
  \nonumber\\
  && 
  - \left( {71\over 144} + {1\over 3}\*\z2 + {13\over 36}\*\lm + {1\over 12}\*\lm^2 \right)\*n_f 
  + {4\over 3}\sum_i\Delta(m_i/\mt)
  \, .
\end{eqnarray}
Here $n_f$ denotes the number of light flavors and $\lm = \ln(\mur^2/\mmu^2)$.
The function $\Delta(m_i/\mt)$ accounts for all massive quarks $m_i$ lighter than the top-quark. 
For all light quarks we set $m_i=0$ so the sum in Eq.~(\ref{eq:d2def}) vanishes.
Note also that the decoupling of the top-quark is assumed 
to be done at the scale of the \msbar\ mass $\mmu$. 

Let us start by making the mass dependence in the total cross section manifest
order by order in perturbation theory. 
For the pole mass $\mt$ we have through NNLO
\begin{equation}
  \label{eq:hadro-mpole}
  \sigma = 
    a_s^2\, \sum_{i=0}^2\, a_s^i\, \sigma^{(i)}(\mt) 
  \, .
\end{equation}
Next, we use the relation~(\ref{eq:mpole-mbar}) above to convert from the pole mass to the \msbar\ mass $\mm$. 
For simplicity we abbreviate $\mbar =\mm $ and obtain
\begin{equation}
  \label{eq:hadro-mbar}
  \sigma = 
    a_s^2\, \sum_{i=0}^2\, a_s^i\, \left(
      \sigma^{(i)}(\mbar) 
     + \mbar\, \sum_{l=1}^i\, d^{(l)} \partial_m \sigma^{(i-l)}(m) \biggr|_{m=\mbar} 
     + \delta_{i,2}\, {1 \over 2}\, \left(\mbar d^{(1)}\right)^2 \partial^2_m \sigma^{(0)}(m) \biggr|_{m=\mbar}
  \right)
  \, .
\end{equation}
We note that the coefficients $d^{(i)}$ have to be evaluated for $\mu_r = \mbar$ 
(corresponding to the scale of $\alpha_s$). 
Thus, the task in Eq.~(\ref{eq:hadro-mbar}) 
amounts to determine the derivatives of the cross sections $\sigma^{(i)}$ with respect to the mass.
To do so in practice we have chosen the following approach. 
For all coefficients $\sigma^{(i)}$ we use the ansatz of Eq.~(\ref{eq:hadrofit}) to parametrize the mass dependence.
More precisely we evaluate the hadronic cross section order by order in perturbation theory 
for a fixed renormalization and factorization scale. 
Then, varying the top-quark mass (in the pole mass scheme) and performing a fit similar 
to what has been discussed before in Eq.~(\ref{eq:hadrofit}) 
we obtain the total cross section in the following form:
\begin{equation}
  \label{eq:hadrofitpole}
  \sigma = a_s^2\, \sum_{i=0}^2\, a_s^i\, \sum_{k=0}^N\, (\mt-m_0)^k\, c_{k}^{(i)}
  \, ,
\end{equation}
where $c_{k}^{(i)}$ denote the (order dependent) fit coefficients.
$N$ is the order of the polynomial in $\mt$ ($N=6$ in Eq.~(\ref{eq:hadrofit})) 
and $m_0$ is our fixed reference mass (taken to be $173\GeV$ in Eq.~(\ref{eq:hadrofit})).
Since all dependence on the pole mass $\mt$ is manifest, 
it is now a straightforward exercise to convert to the \msbar\ mass $\mbar$
and to perform the derivatives in Eq.~(\ref{eq:hadro-mbar}),
\begin{eqnarray}
  \label{eq:hadrofitmsbar}
  \sigma 
&=& a_s^2\, \sum_{i=0}^2\, a_s^i\, 
  \sum_{k=0}^N\, \left( (\mbar-m_0)^k\, c_{k}^{(i)}
    + k\, \mbar\, (\mbar-m_0)^{k-1}\, \sum_{l=1}^i\, d^{(l)}\, c_{k}^{(i-l)}
  \right.
\\
\nonumber
& &\hspace*{8mm}
\left.
     + \delta_{i,2}\, {1 \over 2}\, k\, (k-1)\, \mbar^2\, (\mbar-m_0)^{k-2} \left(d^{(1)}\right)^2 c_{k}^{(0)}\,
  \right)
  \, ,
\\
\nonumber
&\mbarmeq&
a_s^2\, \left( 
  c_{0}^{(0)} 
  + a_s \left\{   c_{0}^{(1)} + \mbar\, d^{(1)}\, c_{1}^{(0)} 
  \right\}
  + a_s^2 \left\{   c_{0}^{(2)} + \mbar\, d^{(1)}\, c_{1}^{(1)} + \mbar\, d^{(2)}\, c_{1}^{(0)} 
     + \mbar^2\, \left(d^{(1)}\right)^2 c_{2}^{(0)}
  \right\}
\right)
  \, .
\end{eqnarray}
If the expansion point $m_0$ is chosen to be the \msbar\ mass $\mbar$, 
Eq.~(\ref{eq:hadrofitmsbar}) simplifies considerably 
and the truncation of the power series in $\mbar$ to first (second) order is exact at NLO (NNLO).
Generally though, for applications, it is of some advantage to keep $m_0$ at a fixed numerical value 
and to rely on the fact, that our ansatz~(\ref{eq:hadrofitpole}) with a polynomial of high enough degree $N$ 
approximates all coefficients $\sigma^{(i)}$ and their first two derivatives sufficiently well.
As discussed below Eq.~(\ref{eq:hadrofit}), the choice $N=6$ achieves per mille accuracy 
in the phenomenologically interesting range.
We have also checked that the choices $m_0=\mbar$ and $m_0 \neq \mbar$ yield the same result.

We stress again, that we have fixed $\mur = \mbar$ in Eq.~(\ref{eq:hadrofitmsbar}).
However, it is also possible to restore the complete renormalization scale
dependence using the well-known relation for the running coupling 
\begin{equation}
  a_s(\mbar) = a_s(\mur)\*\left( 1 
    +  4\pi^2 a_s(\mur)\*L_{\bar R} \* \b0 
    + (4\pi^2)^2 a_s(\mur)^2\* ( \b1\*L_{\bar R} + \b0^2 \*L_{\bar R}^2)\right)
  \, ,
\end{equation}
with $L_{\bar R} = \ln(\mur^2 / \mbar^2)$ and $\b0$ and $\b1$ given in Eq.~(\ref{eq:defbeta}).
To summarize, Eq.~(\ref{eq:hadrofitmsbar}) represents an explicit expression 
for the total cross section of top-quark hadro-production with the top-quark mass defined in
the \msbar\ scheme.

\begin{figure}[ht!]
\centering
\vspace*{10mm}
\subfigure[\label{subfig:msbarscale:left}]
    {
        \includegraphics[bb = 30 55 540 500, scale = 0.385]{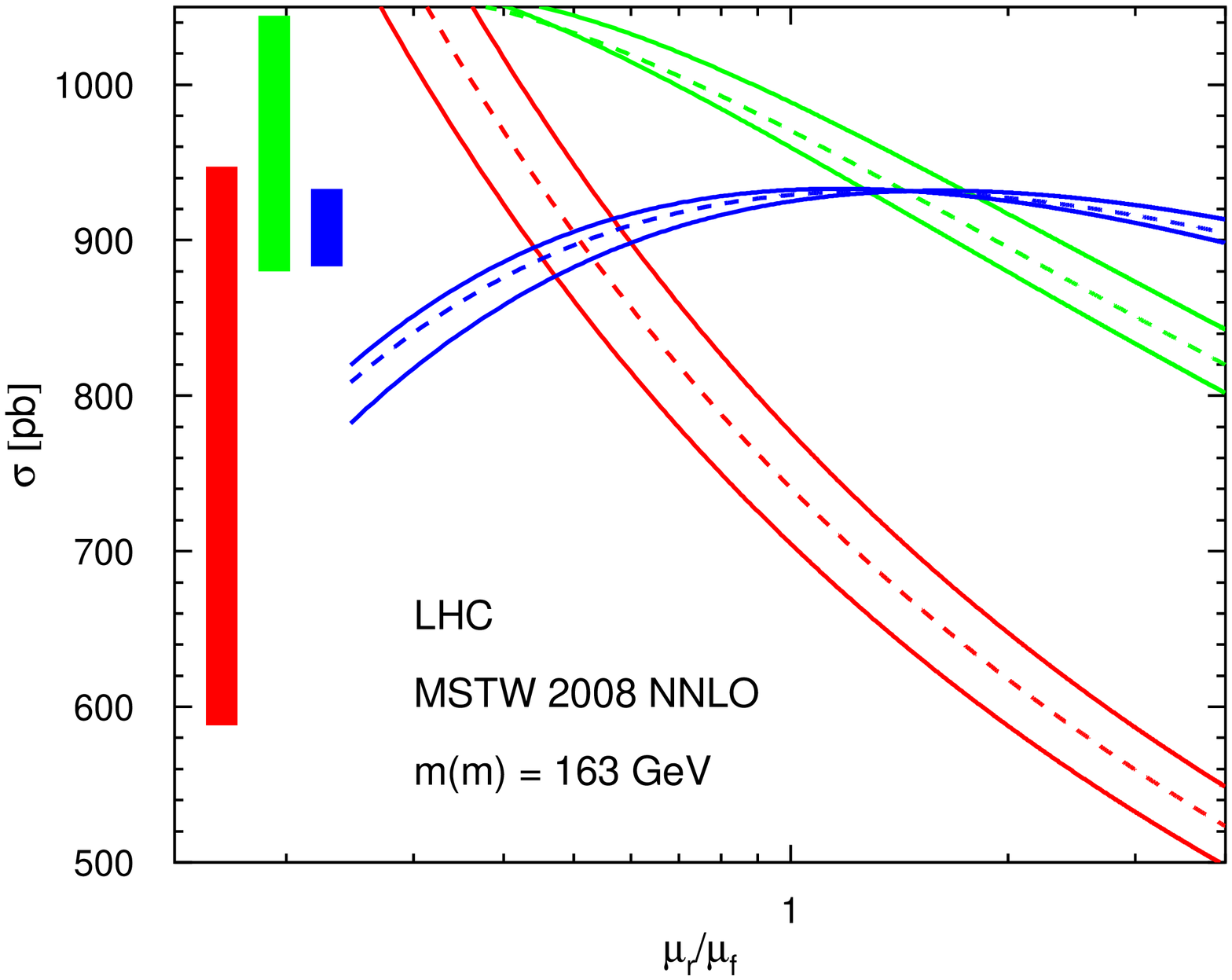}
    }
\hspace*{10mm}
\subfigure[\label{subfig:masbarscale:right}]
    {
        \includegraphics[bb = 30 55 540 500, scale = 0.385]{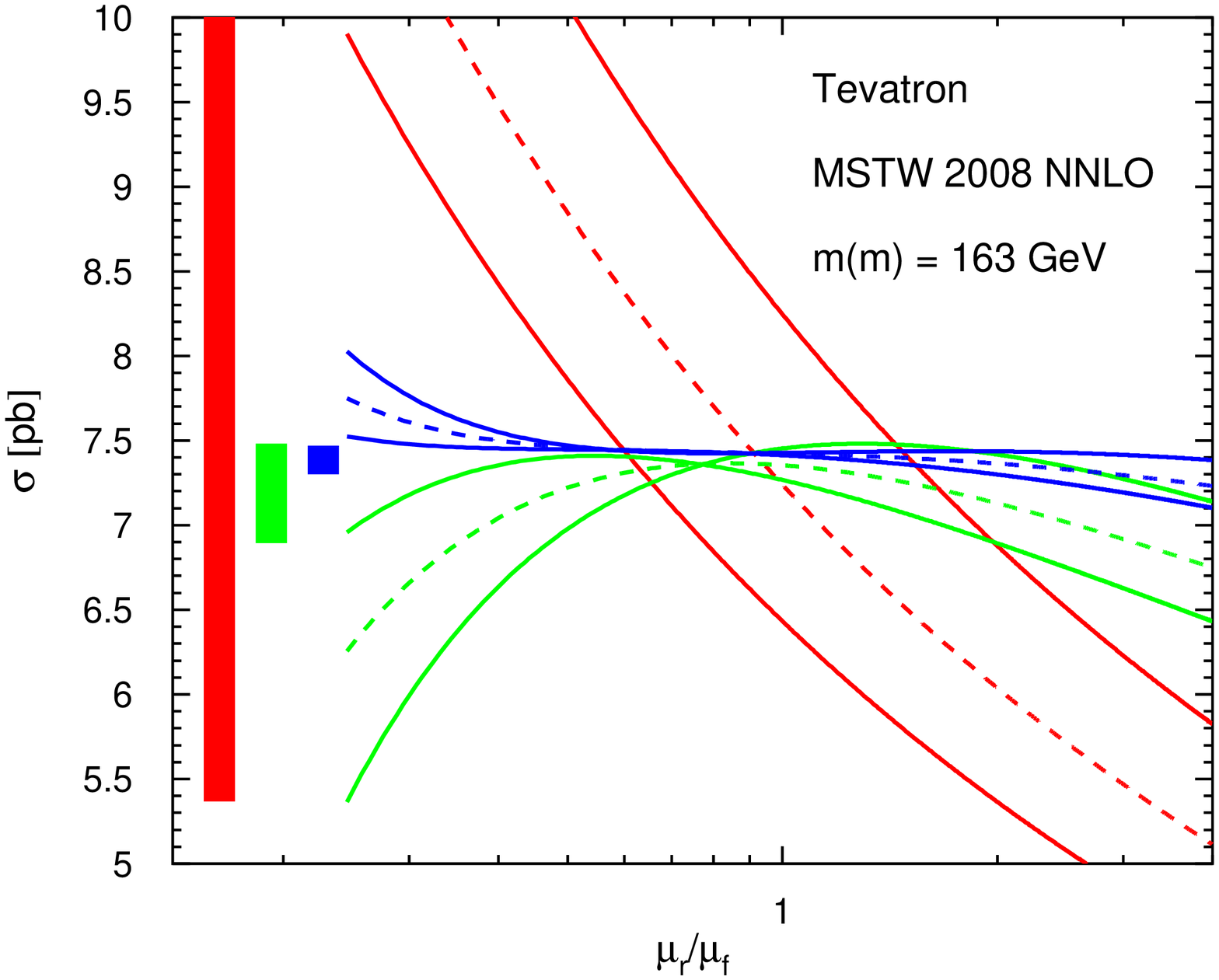}
    }
\hspace*{3mm}
\caption{\small
  \label{fig:msbarscale}
 The scale dependence of the total cross section with the top-quark mass 
 in the \msbar\ scheme at $\mbar = 163\GeV$ at LO (red), NLO (green) and
 approximate NNLO (blue).
 The dashed lines denote the $\muf = \mbar$ for the factorization scale, 
 the solid lines the maximal deviations for $\mur \in [\mbar/2, 2\mbar]$ and $\muf = \mbar/2, \mbar$ and $2\mbar$.
 We use the MSTW 2008 PDF set~\cite{Martin:2009iq} at LHC with $\sqrt{S} = 14\TeV$ (left) 
 and Tevatron with $\sqrt{S} = 1.96\TeV$ (right).
 The vertical bars indicate the size of the scale variation in the standard
 range $[\mbar/2, 2\mbar]$. 
}
\end{figure}
Let us illustrate the phenomenological consequences of the \msbar\ mass for predictions at Tevatron and LHC. 
In Fig.~\ref{fig:msbarscale} we plot the scale dependence of the total cross
section again at the various orders in perturbation theory. 
The value of $\mbar = 163 \GeV$ roughly corresponds to a pole mass of $\mt = 173 \GeV$ 
and we choose three (fixed) values for the factorization scale $\muf = \mbar/2, \mbar$ and $2\mbar$.
The band to the left denotes the maximum and the minimum  values of $\mur \in [\mbar/2, 2\mbar]$ 
for the three choices of $\muf$ according to Eq.~(\ref{eq:range}), 
cf. the contour plot in Fig.~\ref{fig:totalXsection} for the pole mass.
We observe a great stability with respect to scale variations when including higher order perturbative corrections through NNLO.
Remarkably, at Tevatron, the scale variation at NNLO is even reduced further by more than a
factor of two compared to the result in the pole mass scheme.

\begin{figure}[ht!]
\centering
\vspace*{10mm}
\subfigure[\label{subfig:massdep:left}]
    {
        \includegraphics[bb = 30 55 540 500, scale = 0.355]{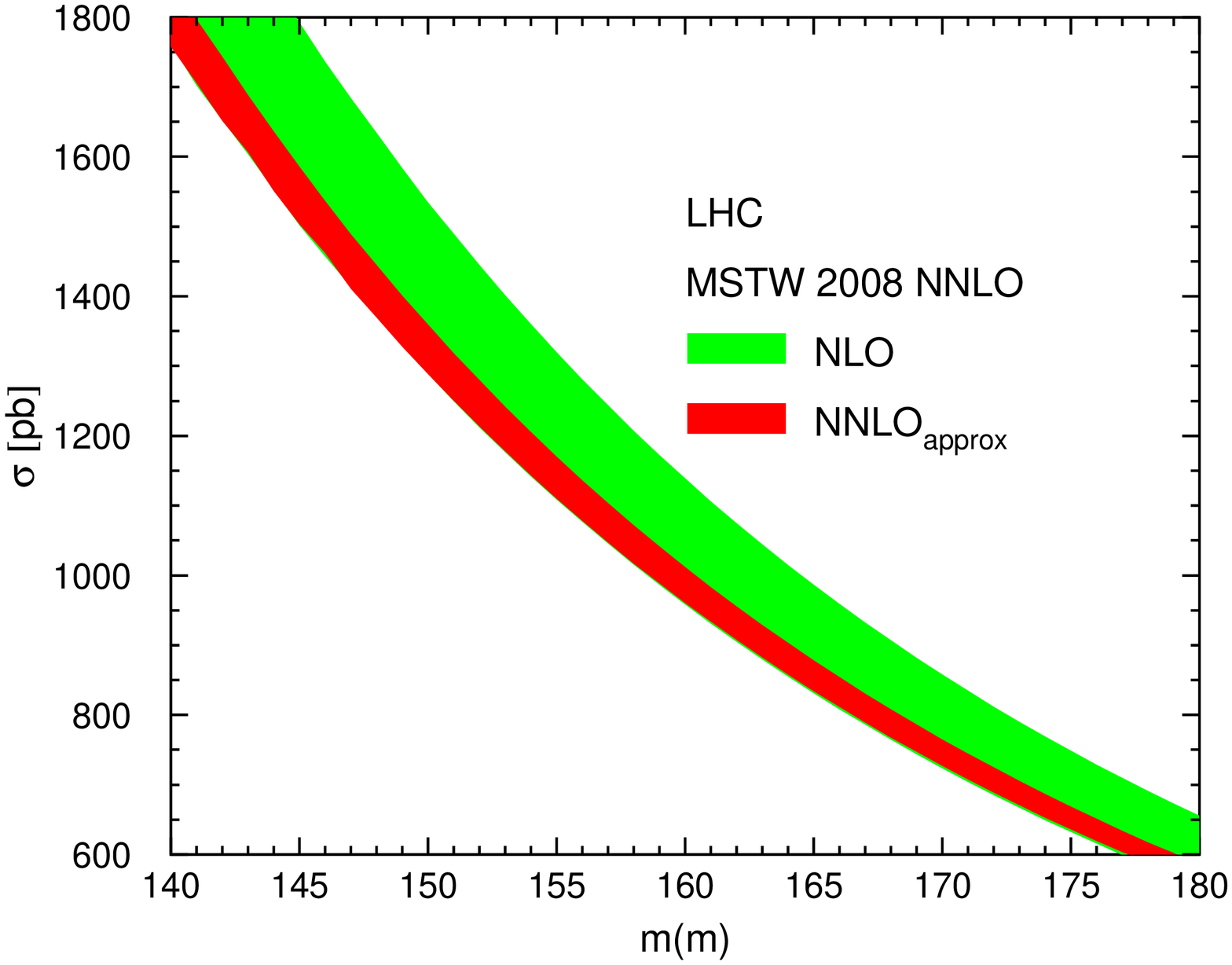}
    }
\hspace*{10mm}
\subfigure[\label{subfig:massdep:right}]
    {
        \includegraphics[bb = 30 55 540 500, scale = 0.355]{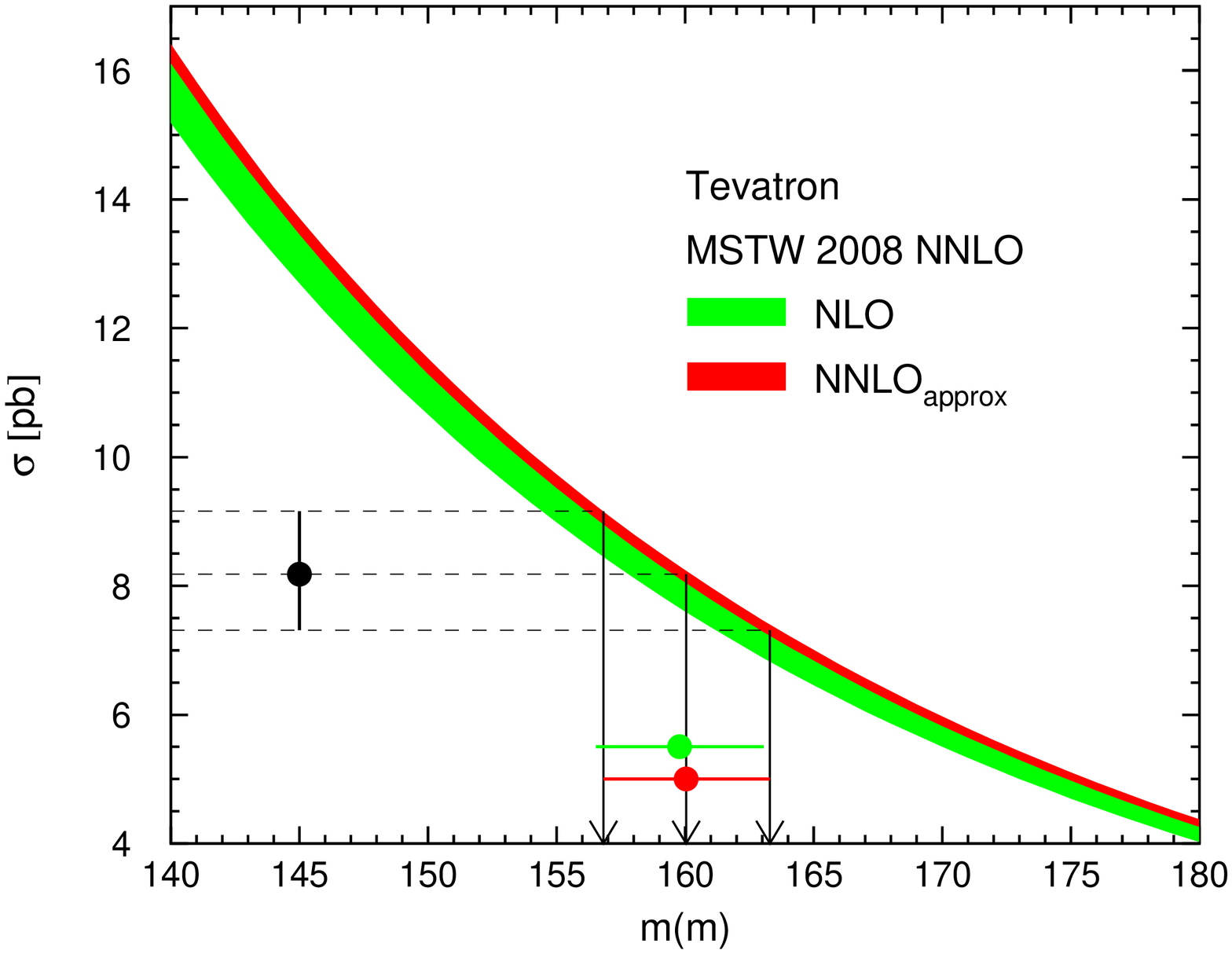}
    }
\hspace*{3mm}
\caption{\small
  \label{fig:massdep}
  The mass dependence of the total cross section for the \msbar\ mass $\mbar$ 
  at NLO (green) and approximate NNLO (red) order with 
  the scale variation in the range $\mur \in [\mbar/2,2\mbar]$
  and $\muf = \mbar/2, \mbar$ and $2\mbar$  
  for the MSTW 2008 PDF set~\cite{Martin:2009iq} at LHC with $\sqrt{S} = 14\TeV$ (left) 
  and Tevatron with $\sqrt{S} = 1.96\TeV$ (right).
  The value for the Tevatron cross section is taken from Ref.~\cite{Abazov:2009ae}.
}
\end{figure}
Next, in Fig.~\ref{fig:massdep} we show the mass dependence of the total cross section 
employing the \msbar\ mass definition and performing the same scale variation
as above, i.e. $\muf = \mbar/2, \mbar$ and $2\mbar$ and $\mur \in [\mbar/2, 2\mbar]$.
Upon adding the higher order perturbative corrections we observe 
as a striking feature the extremely small numerical effect of the radiative corrections.
E.g. for $\mbar = 163\GeV$ at Tevatron, we find the effect of the NLO corrections 
to be only $1.5 \%$ and even less ($0.9 \%$) for the approximate NNLO results. 
Also for the LHC, we observe a much faster convergence of the
perturbative expansion when using the \msbar\ mass. 
The NLO (approximate NNLO) corrections amount to $31 \%$ ($4 \%$) at $\mbar = 163\GeV$ 
which is roughly half of the size of the corrections in the pole mass scheme.
This demonstrates an extremely good stability of the perturbative series in the \msbar\
mass scheme. 
We can understand this behavior qualitatively by looking at the mass dependence 
of the scaling functions in Eqs.~(\ref{eq:fqq0})--(\ref{eq:fgg10}). 
We find e.g. $\partial_m f_{ij}^{(0)} \simeq (1-\beta^2)/\beta$, 
which implies sizably enhanced contributions near partonic threshold, 
i.e. precisely in the region which contributes dominantly in the convolution 
with the parton luminosities, cf. Eq.~(\ref{eq:totalcrs}).
This observation is yet another argument in favor of the phenomenological
importance of our approximate NNLO predictions in Eqs.~(\ref{eq:fqq20})--(\ref{eq:fgg20}).

\begin{table}[ht!]
\begin{center}
\renewcommand{\arraystretch}{1.35}
  \begin{tabular}{l|c|c}
      &  $\mbar$ [$\GeV$] 
      &  $\mt$ [$\GeV$]
      \\ \hline
      LO 
      &$159.2^{+3.5}_{-3.4}$
      &$159.2^{+3.5}_{-3.4}$
      \\
      NLO
      &$159.8^{+3.3}_{-3.3}$
      &$165.8^{+3.5}_{-3.5}$
      \\
      NNLO
      &$160.0^{+3.3}_{-3.2}$
      &$168.2^{+3.6}_{-3.5}$
      \\
      \hline 
  \end{tabular}
\end{center}
\caption{\small
  \label{tab:massnumbers}
 The  LO, NLO and approximate NNLO results for the top-quark mass in the 
 \msbar\ scheme ($\mbar$) and the pole mass scheme ($\mt$) for the 
 measured cross section of $\sigma = 8.18 \pb$ at Tevatron~\cite{Abazov:2009ae}.
 The uncertainties in the Table reflect the quoted experimental uncertainties.
}
\end{table}
A different way to address the issue of perturbative stability is the extraction of the \msbar\ mass from the 
total cross section as measured at Tevatron. 
Ref.~\cite{Abazov:2009ae} quotes a value with a combined uncertainty 
of $\sigma = 8.18^{+0.98}_{-0.87}~\pb$ for a top-quark mass $\mt = 170\GeV$ 
along with a (weak) dependence on the value of the mass, 
e.g. $\sigma = 7.99~\pb$ for the latest world average~\cite{fnal:2009ec}, $\mt = 173.1^{+1.3}_{-1.3} \GeV$.
Using the measured value of $\sigma = 8.18 \pb$ which is consistent with the theory predictions of Refs.~\cite{Moch:2008qy,Moch:2008ai} 
(and with this work, see Tab.~\ref{tab:massnumbers}) we extract the \msbar\ mass $\mbar$ order by order. 
As mentioned earlier we use the same NNLO PDF set of MSTW 2008~\cite{Martin:2009iq} independent the order of perturbation theory 
and the results at LO, NLO and approximate NNLO are given in Tab.~\ref{tab:massnumbers}.
The value of $\mbar=160.0^{+3.3}_{-3.2} \GeV$ represents to the best of our knowledge the 
first direct determination of the running top-quark mass from experimental data.
For comparison, we also quote the values of the pole mass $\mt$ at the
respective order 
extracted in the same way by directly comparing the theory prediction with the measured cross section.
Alternatively, we can also convert the \msbar\ mass value back to the
pole mass scheme with the help of Eq.~(\ref{eq:mpole-mbar}). 
Our NNLO value for $\mbar$ corresponds to $\mt = 168.9^{+3.5}_{-3.4} \GeV$, 
which constitutes a theoretically well-defined determination of the pole mass
and is also (within the experimental uncertainties) in agreement with the world average~\cite{fnal:2009ec} of $\mt = 173.1^{+1.3}_{-1.3} \GeV$.
To summarize, 
the \msbar\ mass scheme is distinguished by the great stability in the value of the extracted top-quark mass.
This feature has been studied in the past in detail for processes at a future linear collider~\cite{Hoang:2000yr} and 
our observation is also in agreement with recent considerations based 
on the renormalization group flow for heavy quark masses~\cite{Hoang:2008yj}.

\subsection*{Summary}
In this Letter, we update and extend the predictions of Refs.~\cite{Moch:2008qy,Moch:2008ai} 
for the cross section of top-quark hadro-production at LHC and Tevatron.
We have applied some improvements in the threshold approximation
for the two-loop scaling functions~(\ref{eq:fqq20})--(\ref{eq:fgg20}) as described in the text.
We provide new and precise parametrizations in Eqs.~(\ref{eq:fqq10})--(\ref{eq:fgg10}) 
and (\ref{eq:fqq21})--(\ref{eq:fgg22}) for all scaling functions 
that can be determined exactly. 
All fit functions are documented in the Appendix.
Moreover, we have performed the independent variation 
of the renormalization and the factorization scale with the help 
of Eqs.~(\ref{eq:scaling-Rqq}), (\ref{eq:scaling-Rgq}).
As a novel aspect, in addition to the conventionally used pole mass 
we provide predictions for the total cross section 
employing the \msbar\ definition for the mass parameter. 
The central result is Eq.~(\ref{eq:hadrofitmsbar}).

Our main phenomenological results are the parametrizations in Eq.~(\ref{eq:hadrofit}), 
Tabs.~\ref{tab:hadrofitcteq} and \ref{tab:hadrofitmstw} 
and the cross sections in Eqs.~(\ref{eq:lhcnumber})--(\ref{eq:cteqtevnumber}) for
the pole mass. 
The differences with respect to our previous numbers are quite small,
though, see Tab.~\ref{tab:crsnumbers}.
The theory uncertainty according to Eq.~(\ref{eq:range}) 
defined by exploring the $(\mur,\muf)$-plane in the standard range $\mur,\muf \in [\mt/2,2\mt]$ 
does not differ significantly from the case of fixed scales $\mur = \muf$.
We have also addressed the residual systematical uncertainty due to the threshold approximation 
and we have quantified the effect of other higher
order corrections, such as electro-weak or QCD bound state effects.
The most interesting aspect of our phenomenological studies consists 
of the conversion to the \msbar\ mass scheme in Figs.~\ref{fig:msbarscale} and \ref{fig:massdep}.
The cross section predictions with the \msbar\ mass definition
exhibit a greatly improved pattern of apparent convergence for the
perturbative expansion and very good stability with respect to scale variations.
This leads to very stable values for the extracted mass parameter $\mbar$ as given in Tab.~\ref{tab:massnumbers}.
In particular we find
\begin{equation}
\label{eq:msbarmass}
\mbar=160.0^{+3.3}_{-3.2} \GeV 
\, .
\end{equation}
This is the first direct determination of the running top-quark mass from top-quark pair-production.
The corresponding value for the pole mass derived from Eq.~(\ref{eq:msbarmass}) reads
\begin{equation}
\mt = 168.9^{+3.5}_{-3.4} \GeV
\, ,
\end{equation}
which is consistent with current world average~\cite{fnal:2009ec}, $\mt = 173.1^{+1.3}_{-1.3} \GeV$.
Altogether, this provides substantial support in view of the reliability of our approximate NNLO numbers.
We believe that the QCD radiative correction for top-quark pair-production at
hadron colliders are well under control.

\begin{sidewaystable}[h!]
\renewcommand{\arraystretch}{1.25}
{\small
\centering
\begin{tabular}{l|rrrrrrr}
\hline
&\multicolumn{1}{c}{$a[\pb]$}&\multicolumn{1}{c}{$b[\pb]$}&\multicolumn{1}{c}{$c[\pb]$}&
 \multicolumn{1}{c}{$d[\pb]$}&\multicolumn{1}{c}{$e[\pb]$}&\multicolumn{1}{c}{$f[\pb]$}&
 \multicolumn{1}{c}{$g[\pb]$}\\[1mm]
\hline
\multicolumn{8}{l}{LHC $\sqrt{s} = 14\TeV$, CTEQ6.6}\\
\hline
    $\sigma(\mu = m_t)$ &$    8.74428  \times 10^{  2 } $ &$   -2.35192  \times 10^{  1 } $ &$    3.74083  \times 10^{ -1 } $ &$   -4.62418  \times 10^{ -3 } $ &$    4.99329  \times 10^{ -5 } $ &$   -4.55463  \times 10^{ -7 } $ &$    2.37050  \times 10^{ -9 } $  \\[1mm]
  $\sigma(\mu = m_t/2)$ &$    8.72517  \times 10^{  2 } $ &$   -2.34260  \times 10^{  1 } $ &$    3.72103  \times 10^{ -1 } $ &$   -4.59525  \times 10^{ -3 } $ &$    4.95790  \times 10^{ -5 } $ &$   -4.51713  \times 10^{ -7 } $ &$    2.34771  \times 10^{ -9 } $  \\[1mm]
  $\sigma(\mu = 2 m_t)$ &$    8.41176  \times 10^{  2 } $ &$   -2.26414  \times 10^{  1 } $ &$    3.60329  \times 10^{ -1 } $ &$   -4.45624  \times 10^{ -3 } $ &$    4.81468  \times 10^{ -5 } $ &$   -4.39515  \times 10^{ -7 } $ &$    2.28915  \times 10^{ -9 } $  \\[1mm]
   $\sigma(\mu = m_t) + \Delta \sigma$ &$    9.02378  \times 10^{  2 } $ &$   -2.40942  \times 10^{  1 } $ &$    3.81862  \times 10^{ -1 } $ &$   -4.71806  \times 10^{ -3 } $ &$    5.10568  \times 10^{ -5 } $ &$   -4.67281  \times 10^{ -7 } $ &$    2.43820  \times 10^{ -9 } $  \\[1mm]
    $\sigma(\mu = m_t) - \Delta \sigma$&$    8.46479  \times 10^{  2 } $ &$   -2.29441  \times 10^{  1 } $ &$    3.66298  \times 10^{ -1 } $ &$   -4.53009  \times 10^{ -3 } $ &$    4.88109  \times 10^{ -5 } $ &$   -4.43754  \times 10^{ -7 } $ &$    2.30362  \times 10^{ -9 } $  \\[1mm]
\hline
\multicolumn{8}{l}{LHC $\sqrt{s} = 10\TeV$, CTEQ6.6}\\
\hline
    $\sigma(\mu = m_t)$ &$    3.96877  \times 10^{  2 } $ &$   -1.12077  \times 10^{  1 } $ &$    1.85352  \times 10^{ -1 } $ &$   -2.36659  \times 10^{ -3 } $ &$    2.62800  \times 10^{ -5 } $ &$   -2.44841  \times 10^{ -7 } $ &$    1.28959  \times 10^{ -9 } $  \\[1mm]
  $\sigma(\mu = m_t/2)$ &$    3.97124  \times 10^{  2 } $ &$   -1.11889  \times 10^{  1 } $ &$    1.84706  \times 10^{ -1 } $ &$   -2.35501  \times 10^{ -3 } $ &$    2.61183  \times 10^{ -5 } $ &$   -2.42989  \times 10^{ -7 } $ &$    1.27805  \times 10^{ -9 } $  \\[1mm]
  $\sigma(\mu = 2 m_t)$ &$    3.79852  \times 10^{  2 } $ &$   -1.07358  \times 10^{  1 } $ &$    1.77667  \times 10^{ -1 } $ &$   -2.26977  \times 10^{ -3 } $ &$    2.52223  \times 10^{ -5 } $ &$   -2.35220  \times 10^{ -7 } $ &$    1.24016  \times 10^{ -9 } $  \\[1mm]
    $\sigma(\mu = m_t)+ \Delta \sigma$ &$    4.15125  \times 10^{  2 } $ &$   -1.15947  \times 10^{  1 } $ &$    1.90285  \times 10^{ -1 } $ &$   -2.41772  \times 10^{ -3 } $ &$    2.67843  \times 10^{ -5 } $ &$   -2.49480  \times 10^{ -7 } $ &$    1.31488  \times 10^{ -9 } $  \\[1mm]
    $\sigma(\mu = m_t)- \Delta \sigma$ &$    3.78628  \times 10^{  2 } $ &$   -1.08207  \times 10^{  1 } $ &$    1.80416  \times 10^{ -1 } $ &$   -2.31532  \times 10^{ -3 } $ &$    2.57769  \times 10^{ -5 } $ &$   -2.40323  \times 10^{ -7 } $ &$    1.26569  \times 10^{ -9 } $  \\[1mm]
\hline
\multicolumn{8}{l}{Tevatron $\sqrt{s} = 1.96\TeV$, CTEQ6.6}\\
\hline
    $\sigma(\mu = m_t)$ &$    7.34317  \times 10^{  0 } $ &$   -2.27486  \times 10^{ -1 } $ &$    3.94086  \times 10^{ -3 } $ &$   -5.22302  \times 10^{ -5 } $ &$    6.09497  \times 10^{ -7 } $ &$   -5.99414  \times 10^{ -9 } $ &$    3.27925  \times 10^{ -11 } $  \\[1mm]
  $\sigma(\mu = m_t/2)$ &$    7.58312  \times 10^{  0 } $ &$   -2.34571  \times 10^{ -1 } $ &$    4.05822  \times 10^{ -3 } $ &$   -5.37018  \times 10^{ -5 } $ &$    6.25408  \times 10^{ -7 } $ &$   -6.13901  \times 10^{ -9 } $ &$    3.35467  \times 10^{ -11 } $  \\[1mm]
  $\sigma(\mu = 2 m_t)$ &$    6.96303  \times 10^{  0 } $ &$   -2.15748  \times 10^{ -1 } $ &$    3.73128  \times 10^{ -3 } $ &$   -4.93012  \times 10^{ -5 } $ &$    5.73218  \times 10^{ -7 } $ &$   -5.62092  \times 10^{ -9 } $ &$    3.07038  \times 10^{ -11 } $  \\[1mm]
    $\sigma(\mu = m_t)+ \Delta \sigma$ &$    7.75854  \times 10^{  0 } $ &$   -2.42254  \times 10^{ -1 } $ &$    4.23665  \times 10^{ -3 } $ &$   -5.65955  \times 10^{ -5 } $ &$    6.63296  \times 10^{ -7 } $ &$   -6.52935  \times 10^{ -9 } $ &$    3.57062  \times 10^{ -11 } $  \\[1mm]
    $\sigma(\mu = m_t)- \Delta \sigma$ &$    6.92780  \times 10^{  0 } $ &$   -2.12718  \times 10^{ -1 } $ &$    3.64506  \times 10^{ -3 } $ &$   -4.78628  \times 10^{ -5 } $ &$    5.55679  \times 10^{ -7 } $ &$   -5.46023  \times 10^{ -9 } $ &$    2.99003  \times 10^{ -11 } $  \\[1mm]
\hline
\end{tabular}
\caption{\small
\label{tab:hadrofitcteq}
Fit coefficients to Eq.~(\ref{eq:hadrofit}) for $\sigma(\mu = \mt, 2 \mt, \mt/2)$ and 
       $\sigma(\mu=\mt)\pm \Delta \sigma$ for the PDF set CTEQ6.6~\cite{Nadolsky:2008zw} and the colliders
       LHC and Tevatron.}
}
\end{sidewaystable}

\begin{sidewaystable}[ht!]
\renewcommand{\arraystretch}{1.25}
{\small
\centering
\begin{tabular}{l|rrrrrrr}
\hline
&\multicolumn{1}{c}{$a[\pb]$}&\multicolumn{1}{c}{$b[\pb]$}&\multicolumn{1}{c}{$c[\pb]$}&
 \multicolumn{1}{c}{$d[\pb]$}&\multicolumn{1}{c}{$e[\pb]$}&\multicolumn{1}{c}{$f[\pb]$}&
 \multicolumn{1}{c}{$g[\pb]$}\\[1mm]
\hline
\multicolumn{8}{l}{LHC $\sqrt{s} = 14\TeV$, MSTW 2008 NNLO}\\
\hline
    $\sigma(\mu = m_t)$ &$    8.87496  \times 10^{  2 } $ &$   -2.38344  \times 10^{  1 } $ &$    3.78224  \times 10^{ -1 } $ &$   -4.66307  \times 10^{ -3 } $ &$    5.02155  \times 10^{ -5 } $ &$   -4.56910  \times 10^{ -7 } $ &$    2.37374  \times 10^{ -9 } $  \\[1mm]
  $\sigma(\mu = m_t/2)$ &$    8.85530  \times 10^{  2 } $ &$   -2.37387  \times 10^{  1 } $ &$    3.76203  \times 10^{ -1 } $ &$   -4.63331  \times 10^{ -3 } $ &$    4.98411  \times 10^{ -5 } $ &$   -4.52980  \times 10^{ -7 } $ &$    2.35137  \times 10^{ -9 } $  \\[1mm]
 $\sigma(\mu = 2 m_t) $ &$    8.54052  \times 10^{  2 } $ &$   -2.29566  \times 10^{  1 } $ &$    3.64547  \times 10^{ -1 } $ &$   -4.49661  \times 10^{ -3 } $ &$    4.84539  \times 10^{ -5 } $ &$   -4.41574  \times 10^{ -7 } $ &$    2.29907  \times 10^{ -9 } $  \\[1mm]
   $\sigma(\mu = m_t) + \Delta \sigma$ &$    9.02902  \times 10^{  2 } $ &$   -2.41907  \times 10^{  1 } $ &$    3.83190  \times 10^{ -1 } $ &$   -4.71808  \times 10^{ -3 } $ &$    5.07642  \times 10^{ -5 } $ &$   -4.61804  \times 10^{ -7 } $ &$    2.39989  \times 10^{ -9 } $  \\[1mm]
    $\sigma(\mu = m_t) - \Delta \sigma$&$    8.72090  \times 10^{  2 } $ &$   -2.34783  \times 10^{  1 } $ &$    3.73257  \times 10^{ -1 } $ &$   -4.60776  \times 10^{ -3 } $ &$    4.96661  \times 10^{ -5 } $ &$   -4.52297  \times 10^{ -7 } $ &$    2.35168  \times 10^{ -9 } $  \\[1mm]
\hline
\multicolumn{8}{l}{LHC $\sqrt{s} = 10\TeV$, MSTW 2008 NNLO}\\
\hline
    $\sigma(\mu = m_t)$ &$    4.03219  \times 10^{  2 } $ &$   -1.13904  \times 10^{  1 } $ &$    1.88177  \times 10^{ -1 } $ &$   -2.39835  \times 10^{ -3 } $ &$    2.65811  \times 10^{ -5 } $ &$   -2.47337  \times 10^{ -7 } $ &$    1.30217  \times 10^{ -9 } $  \\[1mm]
  $\sigma(\mu = m_t/2)$ &$    4.03439  \times 10^{  2 } $ &$   -1.13695  \times 10^{  1 } $ &$    1.87488  \times 10^{ -1 } $ &$   -2.38625  \times 10^{ -3 } $ &$    2.64067  \times 10^{ -5 } $ &$   -2.45191  \times 10^{ -7 } $ &$    1.28831  \times 10^{ -9 } $  \\[1mm]
  $\sigma(\mu = 2 m_t) $&$    3.86012  \times 10^{  2 } $ &$   -1.09154  \times 10^{  1 } $ &$    1.80486  \times 10^{ -1 } $ &$   -2.30194  \times 10^{ -3 } $ &$    2.55275  \times 10^{ -5 } $ &$   -2.37710  \times 10^{ -7 } $ &$    1.25272  \times 10^{ -9 } $  \\[1mm]
   $\sigma(\mu = m_t) + \Delta \sigma$ &$    4.11912  \times 10^{  2 } $ &$   -1.16047  \times 10^{  1 } $ &$    1.91287  \times 10^{ -1 } $ &$   -2.43344  \times 10^{ -3 } $ &$    2.69297  \times 10^{ -5 } $ &$   -2.50369  \times 10^{ -7 } $ &$    1.31793  \times 10^{ -9 } $  \\[1mm]
    $\sigma(\mu = m_t) - \Delta \sigma$&$    3.94526  \times 10^{  2 } $ &$   -1.11761  \times 10^{  1 } $ &$    1.85066  \times 10^{ -1 } $ &$   -2.36310  \times 10^{ -3 } $ &$    2.62315  \times 10^{ -5 } $ &$   -2.44419  \times 10^{ -7 } $ &$    1.28819  \times 10^{ -9 } $  \\[1mm]
\hline
\multicolumn{8}{l}{Tevatron $\sqrt{s} = 1.96\TeV$, MSTW 2008 NNLO}\\
\hline
    $\sigma(\mu = m_t)$ &$    7.04217  \times 10^{  0 } $ &$   -2.18800  \times 10^{ -1 } $ &$    3.80366  \times 10^{ -3 } $ &$   -5.06795  \times 10^{ -5 } $ &$    5.96308  \times 10^{ -7 } $ &$   -5.92150  \times 10^{ -9 } $ &$    3.26369  \times 10^{ -11 } $  \\[1mm]
 $\sigma(\mu = m_t/2) $ &$    7.27746  \times 10^{  0 } $ &$   -2.25794  \times 10^{ -1 } $ &$    3.92060  \times 10^{ -3 } $ &$   -5.21816  \times 10^{ -5 } $ &$    6.12661  \times 10^{ -7 } $ &$   -6.05762  \times 10^{ -9 } $ &$    3.32360  \times 10^{ -11 } $  \\[1mm]
  $\sigma(\mu = 2 m_t)$ &$    6.67970  \times 10^{  0 } $ &$   -2.07517  \times 10^{ -1 } $ &$    3.60070  \times 10^{ -3 } $ &$   -4.78294  \times 10^{ -5 } $ &$    5.60141  \times 10^{ -7 } $ &$   -5.52680  \times 10^{ -9 } $ &$    3.02883  \times 10^{ -11 } $  \\[1mm]
   $\sigma(\mu = m_t) + \Delta \sigma$ &$    7.18407  \times 10^{  0 } $ &$   -2.22429  \times 10^{ -1 } $ &$    3.85513  \times 10^{ -3 } $ &$   -5.13171  \times 10^{ -5 } $ &$    6.05482  \times 10^{ -7 } $ &$   -6.04397  \times 10^{ -9 } $ &$    3.34549  \times 10^{ -11 } $  \\[1mm]
    $\sigma(\mu = m_t) - \Delta \sigma$&$    6.90028  \times 10^{  0 } $ &$   -2.15171  \times 10^{ -1 } $ &$    3.75214  \times 10^{ -3 } $ &$   -5.00394  \times 10^{ -5 } $ &$    5.87170  \times 10^{ -7 } $ &$   -5.80214  \times 10^{ -9 } $ &$    3.18553  \times 10^{ -11 } $  \\[1mm]
\hline
\end{tabular}
\caption{\small
\label{tab:hadrofitmstw}
Same as Tab.~\ref{tab:hadrofitcteq} for the PDF set
MSTW2008~\cite{Martin:2009iq} at NNLO.
The PDF uncertainty $\Delta \sigma$ has been obtained with the 
68\% confidence level set.
}
}
\end{sidewaystable}

\subsection*{Acknowledgments}
We would like to thank W.~Bernreuther, M.~Cacciari, A.~Hoang and H.~Kawamura for stimulating discussions 
and A.~Mitov for his very lively contributions at LoopFest VIII and the CERN Theory Institute {\it TOP09}.
This work is supported by the Helmholtz Gemeinschaft under contract VH-NG-105 
and by the Deutsche Forschungsgemeinschaft under contract SFB/TR 9.
P.U. acknowledges the support of the Initiative and
Networking Fund of the Helmholtz Gemeinschaft, contract HA-101
("Physics at the Terascale").

\appendix

\renewcommand{\theequation}{A.\arabic{equation}}
\setcounter{equation}{0}
\subsection*{Useful formulae}
\label{sec:appA}
\begin{eqnarray}
\label{eq:topf21qq}
\fqqto &=& 
  \frac{1}{(16\pi^2)^2} \fqqn \biggl[
  - \frac{8192}{9}\* \ln^3 \beta 
  + \biggl( \frac{12928}{3} - \frac{32768}{9}\* \ln 2 \biggr)\* \ln^2 \beta
\\
&& 
  + \biggl( - 840.51065 + 70.183854 \frac{1}{\beta} \biggr)\* \ln \beta
  - 82.246703\frac{1}{\beta} + 467.90402 
  \biggr] 
\nonumber\\
&&
  + \frac{n_f}{(16\pi^2)^2} \fqqn \biggl[
  - \frac{256}{3}\* \ln^2 \beta
  + \biggl( \frac{2608}{9} - \frac{2816}{9}\* \ln 2 \biggr) \* \ln \beta
  + 6.5797363\frac{1}{\beta} - 64.614276\biggr] 
\nonumber\\
&& 
  + h(\beta,b_i + n_f \*c_i)
  - \frac{4 n_f^2}{(16\pi^2)^2} \fqqn \left[
    \frac{4}{3} \ln 2 - \frac{2}{3} \ln \rho - \frac{10}{9}\right]
\, ,
\nonumber\\
\label{eq:topf22qq}
\fqqtt &=& 
  \frac{1}{(16\pi^2)^2} \fqqn \biggl[
  \frac{2048}{9} \* \ln^2 \beta 
  + \biggl( - \frac{7840}{9} + \frac{4096}{9}\* \ln 2 \biggr) \* \ln \beta  
  + 270.89724
  \biggr]
\\
&& 
  + \frac{n_f}{(16\pi^2)^2} \fqqn \biggl[
    \frac{320}{9}\* \ln \beta 
  - \frac{596}{9} 
  + \frac{320}{9}\* \ln 2
  \biggr] 
  + h(\beta,b_i + n_f \*c_i)
  + \frac{4 n_f^2}{3 (16\pi^2)^2} \fqqn 
\, ,
\nonumber\\
\label{eq:topf21gq}
\fgqto &=&
  - \frac{\pi}{(16\pi^2)^2} \beta^3 \biggl[ 
    \frac{770}{27} \ln^2 \beta
  + \biggl( - \frac{6805}{81} + \frac{6160}{81} \ln 2 \biggr) \ln \beta 
  + 0.13707784\frac{1}{\beta} 
\\
&& 
  + 0.22068868 
  \biggr]
- \frac{\pi n_f}{81(16\pi^2)^2} \beta^3 \biggl[
  46 \ln \beta - \frac{163}{3} + 76 \ln 2 
  \biggr]
  + h_{gq}^{(b)}(\beta,b_i + n_f \*c_i)
\nonumber\\
\label{eq:topf22gq}
\fgqtt &=&
  \frac{\pi}{(16\pi^2)^2} \beta^3 \biggl[
  \frac{385}{81} \ln \beta - \frac{1540}{243} + \frac{385}{81} \ln 2 
  \biggr] 
  + h_{gq}^{(b)}(\beta,b_i + n_f \*c_i)
\, ,
\\
\label{eq:topf21gg}
\fggto &=& 
  \frac{1}{(16\pi^2)^2} \fggn \biggl[
  - 4608\* \ln^3 \beta 
  + \biggl(\frac{109920}{7}-18432 \* \ln 2 \biggr)\* \ln^2 \beta
\\
&&
  + \biggl( 69.647185 - 248.15005 \frac{1}{\beta} \biggr) \* \ln \beta
  + 56.867721 \frac{1}{\beta} + 17.010070
  \biggr] 
\nonumber\\ 
&&
  + \frac{n_f}{(16\pi^2)^2} \fggn \biggl[
  - 64 \* \ln^2 \beta
  + \biggl( \frac{4048}{21} - 192 \* \ln 2 \biggr)\* \ln \beta
  - 3.4465285 \frac{1}{\beta} - 37.602004\biggr]
\nonumber\\
&& 
  + h(\beta,b_i + n_f \*c_i)
  \, ,
\nonumber\\
\label{eq:topf22gg}
\fggtt &=& 
  \frac{1}{(16\pi^2)^2} \fggn \biggl[
    1152 \* \ln^2 \beta 
  + ( - 2568 + 2304 \* \ln 2 )\* \ln \beta 
  - 79.74312140 
  \biggr]
\\
&&
  + \frac{n_f}{(16\pi^2)^2} \fggn \biggl[
    16\* \ln \beta 
  - 16 + 16\* \ln 2
  \biggr] 
  + h(\beta,b_i + n_f \*c_i)
\, ,
\nonumber
\end{eqnarray}
where all threshold logarithms $\ln(\beta)$ and the Coulomb corrections ($\sim 1/\beta$) are exact.
The fit functions are given in Eqs.~(\ref{eq:tophgg})--(\ref{eq:tophgqb}) and
all parameters of the fit are listed in Tabs.~\ref{tab:qqfit}--\ref{tab:ggfit}.
The fits to the scaling functions $f_{ij}^{(21)}$, $f_{ij}^{(22)}$ in Eqs.~(\ref{eq:topf21qq})--(\ref{eq:topf22gg}) 
are, in general, accurate at the per mille level. 
Exceptions are regions close to zero, which is not surprising. 
There we retain an accuracy better than one percent.

Fortran subroutines with the parametrizations of all scaling functions
and the coefficient in Tabs.~\ref{tab:qqfit}--\ref{tab:ggfit} are available
from the authors upon request.
\begin{eqnarray}
\label{eq:tophgg}
h(\beta,a_1,\ldots,a_{17})&=& 
    a_{{1}}{\beta}^{2}
  + a_{{2}}{\beta}^{3}
  + a_{{3}}{\beta}^{4}
  + a_{{4}}{\beta}^{5}
\\
&&
  + a_{{5}}{\beta}^{2} \ln \beta 
  + a_{{6}}{\beta}^{3} \ln \beta 
  + a_{{7}}{\beta}^{4} \ln \beta 
  + a_{{8}}{\beta}^{5} \ln \beta 
\nonumber\\
&&  
  + a_{{9}}{\beta}^{2}  \ln^{2} \beta
  + a_{{10}}{\beta}^{3} \ln^{2} \beta
  + a_{{11}}\beta\, \ln \rho 
  + a_{{12}}\beta\,  \ln^{2} \rho
  + a_{{13}}{\beta}^{2} \ln \rho
\nonumber\\
&&  
  + a_{{14}}{\beta}^{2}  \ln^{2} \rho
  + a_{{15}}{\beta}^{3} \ln \rho 
  + a_{{16}}{\beta}^{3}  \ln^{2} \rho
  + a_{{17}}{\beta}^{4} \ln \rho 
\, ,
\nonumber\\
\label{eq:tophgqa}
h_{gq}^{(a)}(\beta,a_1,\ldots,a_{15}) &=&
    a_{{1}}{\beta}^{4}
  + a_{{2}}{\beta}^{5}
  + a_{{3}}{\beta}^{6}
\\
&&
  + a_{{4}}{\beta}^{4} \ln \beta 
  + a_{{5}}{\beta}^{5} \ln \beta 
  + a_{{6}}{\beta}^{6} \ln \beta 
\nonumber\\
&&
  + a_{{7}}{\beta}^{2}\rho\, \ln \rho 
  + a_{{8}}{\beta}^{2}\rho\,  \ln^2 \rho
  + a_{{9}}{\beta}^{3}\rho\, \ln \rho
\nonumber\\
&&
  + a_{{10}}{\beta}^{3}\rho\, \ln^2 \rho
  + a_{{11}}{\beta}^{4}\rho\, \ln \rho 
\nonumber\\
&& 
  + a_{{12}}{\beta}^{4}\rho\, \ln^2 \rho
  + a_{{13}}{\beta}^{2}\rho\, \ln^3 \rho
  + a_{{14}}{\beta}^{2}\rho\, \ln^4 \rho
  + a_{{15}}{\beta}^{2}\rho\, \ln^5 \rho
\, ,
\nonumber\\
\label{eq:tophgqb}
h_{gq}^{(b)}(\beta,a_1,\ldots,a_{18})&=&
    a_{{1}}{\beta}^{3}
  + a_{{2}}{\beta}^{4}
  + a_{{3}}{\beta}^{5}
  + a_{{4}}{\beta}^{6}
  + a_{{5}}{\beta}^{7}
\\
&&
   + a_{{6}}{\beta}^{4} \ln \beta
   + a_{{7}}{\beta}^{5} \ln \beta
   + a_{{8}}{\beta}^{6} \ln \beta
   + a_{{9}}{\beta}^{7} \ln \beta
\nonumber\\
&&
   + a_{{10}}{\beta}^{3} \ln \rho
   + a_{{11}}{\beta}^{3} \ln^2 \rho
   + a_{{12}}{\beta}^{4} \ln \rho
   + a_{{13}}{\beta}^{4} \ln^2 \rho
\nonumber\\
&& 
   + a_{{14}}{\beta}^{5} \ln \rho
   + a_{{15}}{\beta}^{5} \ln^2 \rho
   + a_{{16}}{\beta}^{6} \ln \rho
   + a_{{17}}{\beta}^{6} \ln^2 \rho
   + a_{{18}}{\beta}^{7} \ln \rho
\, .
\nonumber
\end{eqnarray}
\begin{table}[ht!]
\centering
\begin{tabular}{r|r|rr|rr}
  &\multicolumn{1}{c|}{$\fqqon$} &\multicolumn{2}{c|}{$\fqqto$} &\multicolumn{2}{c}{$\fqqtt$}\\[1mm]
  $i$&\multicolumn{1}{c|}{$a_i$} & \multicolumn{1}{c}{$b_i$} & \multicolumn{1}{c|}{$c_i$} &
   \multicolumn{1}{c}{$b_i$} & \multicolumn{1}{c}{$c_i$} \\
\hline
$  1 $ & $    0.07120603  $ & $   -0.15388765  $ & $    -0.07960658  $ & $    0.37947056 $ & $   -0.00224114 $  \\[1mm]
$  2 $ & $   -1.27169999  $ & $    4.85226571  $ & $     0.50111294  $ & $   -4.25138041 $ & $    0.02685576 $  \\[1mm]
$  3 $ & $    1.24099536  $ & $   -7.06602840  $ & $    -0.09496432  $ & $    2.91716094 $ & $   -0.01777126 $  \\[1mm]
$  4 $ & $   -0.04050443  $ & $    2.36935255  $ & $    -0.32590203  $ & $    0.94994470 $ & $   -0.00626121 $  \\[1mm]
$  5 $ & $    0.02053737  $ & $   -0.03634651  $ & $    -0.02229012  $ & $    0.10537529 $ & $   -0.00062062 $  \\[1mm]
$  6 $ & $   -0.31763337  $ & $    1.25860837  $ & $     0.23397666  $ & $   -1.69689874 $ & $    0.00980999 $  \\[1mm]
$  7 $ & $   -0.71439686  $ & $    2.75441901  $ & $     0.30223487  $ & $   -2.60977181 $ & $    0.01631175 $  \\[1mm]
$  8 $ & $    0.01170002  $ & $   -1.26571709  $ & $     0.13113818  $ & $   -0.27215567 $ & $    0.00182500 $  \\[1mm]
$  9 $ & $    0.00148918  $ & $   -0.00230536  $ & $    -0.00162603  $ & $    0.00787855 $ & $   -0.00004627 $  \\[1mm]
$ 10 $ & $   -0.14451497  $ & $    0.15633927  $ & $     0.08378465  $ & $   -0.47933827 $ & $    0.00286176 $  \\[1mm]
$ 11 $ & $   -0.13906364  $ & $    1.79535231  $ & $    -0.09147804  $ & $   -0.18217132 $ & $    0.00111459 $  \\[1mm]
$ 12 $ & $    0.01076756  $ & $    0.36960437  $ & $    -0.01581518  $ & $   -0.04067972 $ & $    0.00017425 $  \\[1mm]
$ 13 $ & $    0.49397845  $ & $   -5.45794874  $ & $     0.26834309  $ & $    0.54147194 $ & $   -0.00359593 $  \\[1mm]
$ 14 $ & $   -0.00567381  $ & $   -0.76651636  $ & $     0.03251642  $ & $    0.08404406 $ & $   -0.00035339 $  \\[1mm]
$ 15 $ & $   -0.53741901  $ & $    5.35350436  $ & $    -0.25679483  $ & $   -0.51918414 $ & $    0.00363300 $  \\[1mm]
$ 16 $ & $   -0.00509378  $ & $    0.39690927  $ & $    -0.01670122  $ & $   -0.04336452 $ & $    0.00017915 $  \\[1mm]
$ 17 $ & $    0.18250366  $ & $   -1.68935685  $ & $     0.07993054  $ & $    0.15957988 $ & $   -0.00115164 $  \\[1mm]
\end{tabular}
\small
\caption{\small
\label{tab:qqfit}
Coefficients for fits of the $q\bar{q}$ scaling functions.}
\end{table}
\begin{table}[ht!]
\centering
\begin{tabular}{r|r|rr|rr}
  &\multicolumn{1}{c|}{$\fgqon$} &\multicolumn{2}{c|}{$\fgqto$} &\multicolumn{2}{c}{$\fgqtt$}\\[1mm]
  $i$&\multicolumn{1}{c|}{$a_i$} & \multicolumn{1}{c}{$b_i$} & \multicolumn{1}{c|}{$c_i$} &
   \multicolumn{1}{c}{$b_i$} & \multicolumn{1}{c}{$c_i$} \\
\hline
$  1 $ & $   -0.26103970  $ & $   -0.00120532  $ & $     0.00003257  $ & $   -0.00022247 $ & $    0.00001789 $  \\[1mm]
$  2 $ & $    0.30192672  $ & $   -0.04906353  $ & $     0.00014276  $ & $    0.00050422 $ & $    0.00000071 $  \\[1mm]
$  3 $ & $   -0.01505487  $ & $   -0.20885725  $ & $    -0.00402017  $ & $   -0.02945504 $ & $   -0.00020581 $  \\[1mm]
$  4 $ & $   -0.00142150  $ & $  -13.73137224  $ & $     0.06329831  $ & $    0.34340412 $ & $    0.00108759 $  \\[1mm]
$  5 $ & $   -0.04660699  $ & $   14.01818840  $ & $    -0.05952825  $ & $   -0.31894917 $ & $   -0.00086284 $  \\[1mm]
$  6 $ & $   -0.15089038  $ & $   -0.00930488  $ & $     0.00002694  $ & $    0.00009213 $ & $    0.00000010 $  \\[1mm]
$  7 $ & $   -0.25397761  $ & $   -0.52223668  $ & $     0.00159804  $ & $    0.00690402 $ & $    0.00001638 $  \\[1mm]
$  8 $ & $   -0.00999129  $ & $   -4.68440515  $ & $     0.01522672  $ & $    0.07847233 $ & $    0.00022730 $  \\[1mm]
$  9 $ & $    0.39878717  $ & $   -7.61046166  $ & $     0.02869438  $ & $    0.16042051 $ & $    0.00045698 $  \\[1mm]
$ 10 $ & $   -0.02444172  $ & $    1.36687743  $ & $    -0.00875589  $ & $   -0.05186974 $ & $   -0.00025620 $  \\[1mm]
$ 11 $ & $   -0.14178346  $ & $    1.84698291  $ & $    -0.00800271  $ & $   -0.03861021 $ & $   -0.00016026 $  \\[1mm]
$ 12 $ & $    0.01867287  $ & $   -7.26265988  $ & $     0.04043479  $ & $    0.21650362 $ & $    0.00070713 $  \\[1mm]
$ 13 $ & $    0.00238656  $ & $   -4.89364026  $ & $     0.01965878  $ & $    0.10137656 $ & $    0.00034937 $  \\[1mm]
$ 14 $ & $   -0.00003399  $ & $   11.04566784  $ & $    -0.05262293  $ & $   -0.28056264 $ & $   -0.00072547 $  \\[1mm]
$ 15 $ & $   -0.00000089  $ & $    4.13660190  $ & $    -0.01457395  $ & $   -0.08090469 $ & $   -0.00025525 $  \\[1mm]
$ 16 $ & $    0.00000000  $ & $   -6.33477051  $ & $     0.02314616  $ & $    0.13077889 $ & $    0.00034015 $  \\[1mm]
$ 17 $ & $    0.00000000  $ & $   -1.08995440  $ & $     0.00291792  $ & $    0.01813862 $ & $    0.00006613 $  \\[1mm]
$ 18 $ & $    0.00000000  $ & $    1.19010561  $ & $    -0.00220115  $ & $   -0.01585757 $ & $   -0.00006562 $  \\[1mm]
\end{tabular}
\caption{\small
\label{tab:gqfit}
Coefficients for fits of the $gq$ scaling functions.}
\end{table}
\begin{table}[ht!]
\centering
\begin{tabular}{r|r|rr|rr}
  &\multicolumn{1}{c|}{$\fggon$} &\multicolumn{2}{c|}{$\fggto$} &\multicolumn{2}{c}{$\fggtt$}\\[1mm]
  $i$&\multicolumn{1}{c|}{$a_i$} & \multicolumn{1}{c}{$b_i$} & \multicolumn{1}{c|}{$c_i$} & \multicolumn{1}{c}{$b_i$} & \multicolumn{1}{c}{$c_i$} \\
\hline
$  1 $ & $    -8.92563222  $ & $   -4.18931464  $ & $     0.12306772  $ & $    0.01222783 $ & $   -0.00380386 $  \\[1mm]
$  2 $ & $   149.90572830  $ & $   82.35066406  $ & $    -2.75808806  $ & $   -0.77856184 $ & $    0.08757766 $  \\[1mm]
$  3 $ & $  -140.55601420  $ & $  -87.87311969  $ & $     3.19739272  $ & $    1.33955698 $ & $   -0.10742267 $  \\[1mm]
$  4 $ & $    -0.34115615  $ & $    9.80259328  $ & $    -0.56233045  $ & $   -0.59108409 $ & $    0.02382706 $  \\[1mm]
$  5 $ & $    -2.41049833  $ & $   -1.12268550  $ & $     0.03240048  $ & $    0.00248333 $ & $   -0.00099760 $  \\[1mm]
$  6 $ & $    54.73381889  $ & $   29.51830225  $ & $    -0.92541788  $ & $   -0.23827213 $ & $    0.02932941 $  \\[1mm]
$  7 $ & $    90.91548015  $ & $   48.36110694  $ & $    -1.57154712  $ & $   -0.38868910 $ & $    0.04906147 $  \\[1mm]
$  8 $ & $    -4.88401008  $ & $   -7.06261770  $ & $     0.35109760  $ & $    0.28342153 $ & $   -0.01373734 $  \\[1mm]
$  9 $ & $    -0.17466779  $ & $   -0.08025226  $ & $     0.00227936  $ & $    0.00010876 $ & $   -0.00006986 $  \\[1mm]
$ 10 $ & $    13.47033628  $ & $    7.01493779  $ & $    -0.21030153  $ & $   -0.03383862 $ & $    0.00658371 $  \\[1mm]
$ 11 $ & $    22.66482710  $ & $   15.00588140  $ & $    -0.63688407  $ & $   -0.29071016 $ & $    0.02089321 $  \\[1mm]
$ 12 $ & $     4.60726682  $ & $    3.84142441  $ & $    -0.12959776  $ & $   -0.11473654 $ & $    0.00495414 $  \\[1mm]
$ 13 $ & $   -67.62342328  $ & $  -47.02161789  $ & $     1.91690216  $ & $    0.98929369 $ & $   -0.06553459 $  \\[1mm]
$ 14 $ & $    -9.70391427  $ & $   -8.05583379  $ & $     0.26755747  $ & $    0.24899069 $ & $   -0.01046635 $  \\[1mm]
$ 15 $ & $    65.08050888  $ & $   47.02740535  $ & $    -1.86154423  $ & $   -1.06096321 $ & $    0.06559130 $  \\[1mm]
$ 16 $ & $     5.09663260  $ & $    4.21438052  $ & $    -0.13795865  $ & $   -0.13425338 $ & $    0.00551218 $  \\[1mm]
$ 17 $ & $   -20.12225341  $ & $  -14.99599732  $ & $     0.58155056  $ & $    0.35935660 $ & $   -0.02095059 $  \\[1mm]
\end{tabular}
\caption{\small
\label{tab:ggfit}
Coefficients for fits of the $gg$ scaling functions.}
\end{table}
%

{\footnotesize


}
\end{document}